\newif\ifonecol 
\onecolfalse 

\newif\ifarxiv
\arxivtrue

\ifarxiv
\onecoltrue
\else
\fi

\ifonecol
\documentclass[12pt,draftcls,letterpaper,onecolumn]{IEEEtran}
\else
\documentclass[journal,a4paper]{IEEEtran}
\fi

\newlength{\figurewidth}
\ifonecol
\else
\fi

\ifonecol
\usepackage{setspace}
\fi

\normalsize

\ifCLASSINFOpdf
\else
\fi

\usepackage[tight,footnotesize]{subfigure}

\usepackage{times,amsfonts,amssymb,amsmath,setspace}
\usepackage{color}
\usepackage{graphicx,multirow}
\usepackage{cite}
\usepackage{url}
\usepackage{bbm}\usepackage{mathrsfs}\usepackage{float}
\usepackage{lipsum}
\usepackage{multicol}
\usepackage{caption}
\usepackage{ifpdf}
\ifpdf
  \usepackage{epstopdf}
\fi

\usepackage{graphicx,url,bm,subfigure,comment}

\DeclareMathAlphabet\mathbfcal{OMS}{cmsy}{b}{n}

\newcommand{\pf}{\noindent{\bf Proof:~}}
\newcommand{\qedsymb}{\hfill{\rule{2mm}{2mm}}}

\DeclareMathAlphabet{\mathsfbf}{OT1}{cmss}{sbc}{n}

\newtheorem{lemma}{Lemma}[section]

\newtheorem{proposition}{Proposition}[section]

\newcommand{\EE}{\mathbb{E}} 
\newcommand{\PP}{\mathbb{P}} 
\newcommand{\RR}{\mathbb{R}} 

\newcommand{\ee}{{\rm e}}
\newcommand{\dd}{{\rm\,d}} 

\newcommand{\bv}{{\bf b}}

\newcommand{\dv}{{\bf d}}

\newcommand{\rv}{{\bf r}}
\newcommand{\sv}{{\bf s}}

\newcommand{\uv}{{\bf u}}
\newcommand{\wv}{{\bf w}}
\newcommand{\vv}{{\bf v}}
\newcommand{\xv}{{\bf x}}

\newcommand{\Am}{{\bf A}}
\newcommand{\Bm}{{\bf B}}

\newcommand{\Dc}{{\cal D}}
\newcommand{\Ec}{{\cal E}}

\newcommand{\Ic}{{\cal I}}

\newcommand{\Mc}{{\cal M}}

\newcommand{\Qc}{{\cal Q}}

\def\ben{\begin{enumerate}}
\def\beq{\begin{equation}}
\def\beqa{\begin{eqnarray}}
\def\bit{\begin{itemize}}
\def\een{\end{enumerate}}
\def\eeq{\end{equation}}
\def\eeqa{\end{eqnarray}}
\def\eit{\end{itemize}}

\def\non{\nonumber\\}

\newcommand{\algo}[2]{
\begin{center}
\parbox{0.9\columnwidth}{
\rule{0.9\columnwidth}{0.5mm}\\
\noindent {\bf Algorithm~#1:} 
#2
\rule{0.9\columnwidth}{0.5mm}
}
\end{center}
}

\newtheorem{remark}{\bf Remark}

\usepackage{enumerate}

\usepackage{soul}

\soulregister\cite7
\soulregister\ref7
\soulregister\pageref7

\begin{document}

%
\title{Chaos-Based Anytime Reliable Coded Communications}
%
%
%


\author{Alberto Tarable,~\IEEEmembership{Member,~IEEE,} and Francisco J. Escribano,~\IEEEmembership{Senior Member,~IEEE}%
\thanks{Alberto Tarable is with the Consiglio Nazionale delle Ricerche, Istituto di Elettronica e di Ingegneria Informatica e delle Telecomunicazioni (CNR-IEIIT), Italy (e-mail: alberto.tarable@ieiit.cnr.it).}%
\thanks{ Francisco J. Escribano is with the Department of Signal Theory and Communications, Universidad de Alcal\'a, 28805 Alcal\'a de Henares, Spain (e-mail: francisco.escribano@ieee.org).}}

%
%

\markboth{Journal of \LaTeX\ Class Files,~Vol.~6, No.~1, January~2007}%
{Submitted paper}
%



\maketitle

\begin{abstract}
Anytime reliable communication systems are needed in contexts where the property of vanishing error probability with time is critical. This is the case of unstable real time systems that are to be controlled through the transmission and processing of remotely sensed data. The most successful anytime reliable transmission systems developed so far are based on channel codes and channel coding theory. In this work, another focus is proposed, placing the stress on the waveform level rather than just on the coding level. This alleviates the coding and decoding complexity problems faced by other proposals. To this purpose, chaos theory is successfully exploited in order to design two different anytime reliable alternatives. The anytime reliability property is formally demonstrated in each case for the AWGN channel, under given conditions. The simulation results shown validate the theoretical developments, and demonstrate that these systems can achieve anytime reliability with affordable resource expenditure.
\end{abstract}


%
\IEEEpeerreviewmaketitle

\section{Introduction}

Following a trend that has affected several application fields, the network paradigm has been applied also to automatic control. In particular, it has become more and more important to study the scenario in which the measurement sensor and the controller are not physically co-located, and communicate through a wireless channel. In such a scenario, it is natural to suppose that the controller receives noisy versions of the measurements transmitted by the remote sensor. Channel encoding may therefore be required in order to defend against communication errors.  

From the information-theoretic point-of-view, however, the problem of control through noisy communication channels is rather different from the ordinary reliability problem of a point-to-point link. First of all, the fact that the measurements arrive regularly in time imply that a form of continuous and causal form of encoding-decoding must be sought after, since control must be applied in real time, or within a maximum tolerable delay, so that coding/interleaving solutions that apply to long data batches are not viable in this context. Second, the specific application calls for a coding solution that is able sooner or later to correct every possible past decoding error, to avoid that one of such errors has a catastrophic effect on the evolution of the controlled system at hand. This last observation implies that the coding schemes to be employed for such application satisfy specific performance criteria.

The previous considerations led Sahai and Mitter~\cite{Sahai} to introduce the new concept of \emph{anytime reliability}. Loosely speaking, an encoding-decoding scheme is said to be anytime reliable if its bit error probability decreases exponentially with the decoding delay $d$, i.e., it goes down as $\ee^{-\beta d}$, where $\beta>0$ is the \emph{anytime exponent} of the scheme. Anytime reliable nonlinear tree codes were first proven to exist in \cite{Schulman} and then further developed in \cite{Ostrovsky}. Random linear codes were first introduced in \cite{Como}. Later, Sukhavasi and Hassibi \cite{Hassibi2} showed that causal random linear codes with maximum-likelihood
(ML) decoding are anytime reliable with high probability. Unfortunately, such schemes are characterized by a high decoder complexity, although in \cite{Hassibi2} a decoder with reasonable complexity is proposed for the binary erasure channel (BEC).  Papers~\cite{Dossel, Grosjean} proposed and studied a protograph-based low-density parity-check (LDPC)
convolutional scheme which is shown to achieve anytime reliability on the BEC at an affordable complexity. Such scheme was then proved in~\cite{Tarable} to be anytime reliable also on the binary-input AWGN (BIAWGN) channel. Other anytime reliable coding schemes introduced in the literature, whose properties are shown both on the BEC and on the BIAWGN channel, are based on spatially-coupled LDPC codes~\cite{Noor-A-Rahim, Zhang1} and on repeat-accumulate codes~\cite{Zhang2}. {\ifarxiv
\else
\color{red}
\fi
Recent papers \cite{Grosjean16, Noor-A-Rahim18} put the stress on the availability of a feedback channel from the decoder to the encoder. In both cited papers, it is assumed that the encoder changes according to the received feedback, to improve the error floor that spoils the anytime reliability property of the schemes. In \cite{Grosjean16}, which proposes an LDPC convolutional code, the feedback is used to vary the instantaneous rate of the encoder. In \cite{Noor-A-Rahim18}, which studies a solution based on repeat-accumulate codes, the feedback has the effect of changing the connections in the Tanner graph of the code. See also \cite{GrosjeanPhD} for a comprehensive treatment of the scheme based on LDPC convolutional codes.}   

The basic structure of an anytime reliable coding scheme is well represented by an infinite-memory, ever-growing convolutional trellis. This fact, together with the continuous character of the information to be transmitted (the measurements), allows  thinking that possible solutions for such a scenario could spring from the realm of chaotic systems. Chaotic systems are nonlinear systems that feature some properties, out of which the most important one is the strong sensitivity on the initial conditions. Chaotic systems have been studied for decades and there is now a consolidated general theory that is able to characterize them and that is useful for design and optimization, whenever engineering applications of chaotic systems come at hand.
 
Chaotic waveforms or chaotic encoding possess inherent memory, and are most often derived from discrete-time continuous-amplitude nonlinear maps.
After its introduction in the early 90's, the idea of using chaos in the context of telecommunications has evolved to produce a wide variety of possibilities \cite{Kaddoum16}. Current state-of-the-art shows that chaos-based systems can be usefully employed in typical situations and environments, like wireless communications \cite{Kaddoum17}, multicarrier communications \cite{Yang17}, multiple access \cite{Chen17} or ultra-wideband communications \cite{Mesloub17}. Chaos-based communication systems can work at the waveform or coding levels, or both, like in the case of chaos-based coded modulations (CCM), inspired by the ideas of \cite{Kozic06}. This possibility has produced alternatives that perform well in fading channels \cite{Escribano14}, and has even led to successful practical demonstrations \cite{Wagemakers17}.

{\ifarxiv
\else
\color{red}
\fi
In the present context, the stress has traditionally been put in using variable-memory binary codes of different kinds, which initially adapt well to the demands of variable decoding delay to achieve anytime reliability over noisy channels \cite{Hassibi2,Dossel,Grosjean}. Nevertheless, this requires the subsequent modulation step to create the appropriate waveform for transmission over the corresponding medium. Chaos-based systems, on the contrary, can be driven to perform coded modulation, thus avoiding the usage of two separate steps for the same puspose \cite{Kozic06,Kaddoum16}. On the other hand, chaos-based transmitters and receivers are easy to implement and usually require lower complexity and/or processing delay \cite{Mesloub17}, whereas anytime reliable variable-memory binary codes may not be as efficient in this respect. Moreover, chaos-based waveforms and their nonlinear nature are a better fit to nonlinear transceivers or to nonlinear signal propagation media whose importance is growing nowadays \cite{Grzybowski11} (e.g. free-space optical communications), thus avoiding the usually complex steps required to linearize such communication channels. This opens the scope to deploy hypothetical chaos-based anytime reliable systems in a wider variety of channels, not to mention their potential against dispersive phenomena \cite{Escribano14}. Additionally, chaotic signals have low autocorrelation and may thus provide inherent interference rejection and self-synchronisation capabilities \cite{Chen17}.}
 
To the best of our knowledge, this paper represents the first application of chaos-based communications for anytime reliability. These are the main contributions of the paper:
\begin{itemize}
\item We propose two CCM schemes, where the first has fixed bandwidth efficiency and adaptive instantaneous power, while the second has adaptive bandwidth efficiency and instantaneous power.

\item For both schemes, we derive sufficient conditions for anytime reliability on the AWGN channel, and derive lower bounds to their respective anytime exponents. 

\item Through numerical simulations, we assess the performance of the schemes for different choices of the parameters defining the schemes, and thus obtain useful hints for the design of a practical anytime reliable CCM scheme. 
\end{itemize}

{\ifarxiv
\else
\color{red}
\fi
From the theoretical point of view, it is not mandatory that there exists feedback between receiver and transmitter to achieve anytime reliability \cite{Sahai,Hassibi2}. The chaos-based alternatives proposed in this paper do not inherently require feedback to achieve anytime reliability, and the ensuing mathematical demonstrations will not consider any hypothetical return channel. On the other hand, to render anytime reliable systems usable, it is convenient to have feedback in order to dinamically adjust the parameters of the transmission and keep the required resources within practical limits. As a consequence, and for the purpose of getting illustrative simulation results, the algorithms describing the operation of the proposed chaos-based anytime reliable systems will consider the presence of a return channel, dedicated to communicating the decoding state to the transmitter. We stress that, unlike our approach, for which feedback is needed to meet the practical constraint of limited use of resources such as power and bandwidth, for the schemes in \cite{Grosjean16, Noor-A-Rahim18}, feedback is essential to achieve anytime reliability at finite length.}


The structure of the paper is as follows. In Section~\ref{sec:AR}, we briefly define an anytime reliable system. In Section~\ref{sec:VSCCM}, we describe the first proposed CCM scheme, i.e., adaptive-size CCM, and derive sufficient conditions for its anytime reliability on the AWGN channel. In Section~\ref{sec:UBCCM}, we describe the second proposed CCM scheme, adaptive-bandwidth CCM, and also derive conditions for it to be anytime reliable on the AWGN channel. For both schemes, in Section~\ref{sec:performance}, we show simulation results and give design hints for a practical implementation of the proposed architectures. {\ifarxiv
\else
\color{red}
\fi
Finally, in Section~{\ref{sec:conclusions}}, we draw some conclusions.}

\section{Anytime reliable systems} \label{sec:AR}

Consider a dynamical system that, at discrete time instants, produces vectors of $m$ bits to be transmitted through an AWGN channel. In the following description, we will suppose $m=1$ for simplicity, but the generalization is straightforward. Let $b_n$ be the bit produced by the system at time instant $n$. The system encodes causally all the bits it has produced up to this time into a channel input $s_n$. If $\Ec^{(n)}$ denotes the encoder at time $n$, we have   
%
%
%
\begin{eqnarray*}
s_n &=& \Ec^{(n)}(b_1, \dots, b_n). 
\end{eqnarray*} 

At time $n$, the receiver receives a corrupted version $r_n$ of $s_n$ and  calls a causal decoder $\Dc^{(n)}$, which takes as input all the received symbols and outputs estimates of the past information bits:
\begin{eqnarray*}
(\widehat{b}_1^n,\dots,\widehat{b}_n^n) &=& \Dc^{(n)}(r_1, \dots, r_n). 
\end{eqnarray*} 
where $\widehat{b}_n^{k}$ is the estimate of information bit $b_n$ produced by the decoder at time $k$, with $k\geq n$.

We say that the communication scheme is \emph{anytime reliable} if, for every $n$ and  $d \geq d_0$  
\beq
\PP\{ \widehat{b}_n^{n+d-1} \neq b_n\} < K \ee^{-\gamma d},
\eeq
$K$ and $\gamma$ being positive constants. The infimum of the values of $\gamma$ for which the above condition holds is called the \emph{anytime exponent} of the encoder-decoder pair.

Anytime reliable schemes find their application in control theory, in the scenario when the controller has access only to noisy versions of the system measurements. Consider the discrete-time dynamic time-invariant system
\begin{equation}
\xv_{t+1} = \Am \xv_t + \Bm\uv_t + \vv_t ,
\label{eq:system}
\end{equation}
where $\xv_t\in \RR^{n_x}$ is the state of the system at time step $t$, $\Am$ and $\Bm$ are $n_x \times n_x$ and $n_x \times n_u$ real matrices, respectively, $\uv_t\in \RR^{n_u}$ is the control input, and $\vv_t\in \RR^{n_x}$ is a zero-mean bounded noise process. The system in~\eqref{eq:system} is supposed to be unstable, i.e., it is characterized by $\rho(\Am)>1$, where $\rho(\Am)$ is the spectral radius of the matrix $\Am$, that is the largest magnitude of the eigenvalues of $\Am$.  Sukhavasi and Hassibi derive in \cite{Hassibi2} the conditions under which an anytime reliable encoding-decoding scheme can be used to stabilize the system of \eqref{eq:system} in the mean-square sense, so that the expected value of $\|\mathbf{x}_t\|^2$ is bounded for all $t$.
It is shown in \cite{Hassibi2} that, when using hypercuboidal filters, mean-square sense stability is achieved by a code with anytime exponent $\gamma$ satisfying $\gamma > 2\log\rho(\overline{\Am})$, where $\overline{\Am}$ is the $n_x \times n_x$ matrix whose elements are the absolute values of the elements of $\Am$.

\section{Adaptive-size CCM} \label{sec:VSCCM}

We consider in this paper a time-discrete chaotic map, defined as a nonlinear map\footnote{It is straightforward to generalize the description to a chaotic map defined on a generic interval $\Ic$.} $f: [0,1] \rightarrow [0,1]$. Examples of chaotic maps will be given later in this section. We define the \emph{invariant cdf} $F_f$ of the map as the one which is preserved by the map itself, i.e., if $\xi$ is a R.V. satisfying $\PP \{\xi \leq x\} = F_f(x)$, then also $\PP \{f(\xi) \leq x\} = F_f(x)$. 

We define in the following a CCM scheme able to encode a semiinfinite binary information sequence $\bv = (b_1,b_2,\dots)$. In order to do that, we need two fundamental ingredients:
\begin{itemize}
\item A mapper $\Mc_f$ that maps semiinfinite binary sequences into $[0,1]$, such that the uniform distribution on binary sequences is transformed into the invariant density on $[0,1]$, and

\item A family of (possibly nonuniform) quantizers $(\Qc_f^{(1)},\Qc_f^{(2)},\dots )$ , such that $\Qc_f^{(n)}$ maps a continuous R.V. $\xi$ defined over the interval $[0,1]$ into a set of $Q = 2^n$ equally likely discrete levels. 
\end{itemize}

Notice that the requirement that the mapper output is distributed according to the invariant density is desirable to obtain a system whose statistical properties are stationary. Ideally, given $\bv$, the symbols to be transmitted are obtained by quantization of chaotic samples $z_1,z_2,\dots$ as follows. First, 
\beq
z_n = \left\{
\begin{array}{cc}
\Mc_f(\bv), & n = 1 \\
f^{(\delta_n)}(z_{n-1}), & n > 1
\end{array}
\right. ,
\eeq
where $\delta_n$ is a nonnegative integer whose value will be determined later, and $f^{(\delta_n)}$ means that the chaotic map is applied $\delta_n$ times. Then the quantized chaotic samples are obtained as
\beq
z_n^Q = \Qc_f^{(q_n)}(z_n),
\eeq
where $q_n$ is a positive integer representing the number of quantization bits at time $n$. 
We require the CCM scheme to be causal, so that the quantized sample $z_n^Q$ depends only on the length-$q_n$ subsequence $\bv_{n-q_n+1}^{n}$. In this way, the CCM encoder is practically feasible even if the sequence $\bv$ is not fully known in advance and fits the description of Section \ref{sec:AR}. Let us define a time-shift operator $T$ defined as follows: if $\bv = (b_1,b_2,\dots)$, then $T\bv = (b_2,b_3,\dots)$. The conditions that the mapper must satisfy in order to yield the causality property are the following.
\begin{eqnarray}
f \left(\Mc_f(\bv) \right) & = & \Mc_f\left(T \bv \right), \\
\Qc_f^{(n)} \left(\Mc_f(\bv) \right) & = & \Mc_f\left([\bv_{1}^{n}, 0,0,\dots] \right).
\end{eqnarray}

It is worth noting that, for a map $f$ with nonuniform invariant density, we can always choose $\Mc_f(\bv) = F_f^{-1} \left(\Mc_{f_u}(\bv)\right)$, where $f_u$ is the map with uniform invariant density that is topologically conjugate with $f$. Moreover, we can always choose $\Qc_f^{(n)} = F_f^{-1} \circ \Qc_U^{(n)} \circ F_f$, where $\Qc_U^{(n)}$ is a uniform quantizer on $[0,1]$ with  $2^n$ levels. 

The symbol to be transmitted is denoted $s_n$, which is a scaled and zero-mean version of $z_n^{Q}$. Scaling allows to set the average power, while centering around zero avoids the DC component and makes the scheme more energy-efficient. Let $\rv_1^n$ be the received sequence of samples, which will be supposed to be corrupted by Gaussian noise, i.e.,
\beq
\rv_1^n = \sv_1^n + \wv_1^n,
\eeq
where $\wv_1^n = (w_1,\dots,w_n)$ is a vector of i.i.d. Gaussian noise samples with zero mean and variance $\sigma^2$. The optimal receiver is based on the Viterbi algorithm on a time-varying trellis with $Q_n = 2^{q_n}$ states at time $n$ and outputs the ML estimate of the transmitted chaotic sequence, i.e.,
\beq
\widehat{\sv}_1^n = \arg \max_{\sv_1^n} \PP \{ \rv_1^n | \sv_1^n\}.
\eeq 
A demapper then outputs the estimated bits $\widehat{\bv}_1^n$.

In the following, we make the assumption that the receiver uses a rule to discriminate the bits which are reliably decoded from the bits whose decoding is still unreliable. Suppose that, at time $n$, the oldest (i.e., lowest-indexed) information bit which is not yet reliably decoded is bit $\epsilon_n$. We will also suppose that the receiver is able to convey the value of $\epsilon_n$ to the transmitter through a noise-free dedicated feedback channel. By iterating map $f$ a number of times equal to $\delta_n = \epsilon_n - \epsilon_{n-1}$, the transmitter will discard from the encoder input all bits that are older than $b_{\epsilon_n}$ and use only bits $b_{\epsilon_n},\dots, b_{n+1}$ to generate $z_{n+1}^Q$. Thus, at time $n+1$, $q_{n+1} = n-\epsilon_{n}+2$. 

Algorithm 1 reports the ideal steps of adaptive CCM, as they have been described above. Notice that, if at time $n$ the decoder is not able to decode reliably any of the previously unreliable bits, then $\epsilon_n = \epsilon_{n-1}$, so that the chaotic map is not applied ($\delta_n = 0$) and the number of bits in the quantizer increases by 1. Instead, if all bits up to the current one are found to be reliably decoded, then $\epsilon_n = n+1$ so that the bit queue is emptied and at the next time $q_{n+1} = 1$. 

\begin{table} 
\algo{1}{ 
Adaptive-size CCM.

\begin{itemize}

\item The initial condition is set to $z_1 = \Mc_f \left( \bv \right)$ and  $\epsilon_0 = 1$.

\item Then, for $n = 1, 2, \dots$ 

\begin{enumerate}

\item The transmitter sets $q_n = n - \epsilon_{n-1}+1$. The chaotic sample $z_n$ is quantized over $Q_n = 2^{q_n}$ values by  quantizer $\Qc_f^{(n)}$:
\beq \label{eq:z_n_Q_as}
z_{n}^Q = \Qc_f^{(n)}(z_{n}).
\eeq

\item The quantized chaotic sample is  normalized and centered around zero, i.e.,
\beq \label{eq:s_n_as}
s_n = \Gamma_{q_n} (z_n^Q - \EE\{z_n^Q\}),
\eeq
where $\Gamma_{q_n} $ is a size-dependent normalization constant. Then, $s_n$ is transmitted. 

\item The receiver performs CCM decoding and sends back to the transmitter through the feedback channel the value of $\epsilon_{n}$.

\item $\delta_n = \epsilon_{n}-\epsilon_{n-1}$ steps of the chaotic map are performed, i.e.,
$z_{n+1} = f^{(\delta_{n})}(z_{n})$.
\end{enumerate}

\end{itemize}}
\end{table}

Examples of CCM schemes with different maps are the following, for all of which $\EE\{z_n^Q\} = 1/2$.

\begin{itemize}
\item {\bf Bernoulli shift map (BSM)}: This chaotic map is defined by the recurrence
\beq
\label{BSM}
z_{n+1} = \left\{ 
\begin{array}{cc}
2 z_n, & 0 \leq z_n < 1/2, \\
2 z_n -1, & 1/2 \leq z_n \leq 1.
\end{array}
\right.
\eeq 
It is easily seen that the invariant pdf is the uniform one, so that $F_{\mathrm{BSM}}(x) = x$. Moreover, the binary sequence is mapped into $[0,1]$ according to   $\Mc_{\mathrm{BSM}}(\bv) = \sum_{n=1}^{\infty} b_n 2^{-n}$, i.e., $\bv$ is interpreted as the binary expansion of the chaotic sample.

\item {\bf Tent map}: It is defined by the recurrence 
\beq
z_{n+1} = 1-|2z_n-1|.
\eeq
As in the BSM case, the invariant pdf is the uniform one, so that $F_{\mathrm{Tent}}(x) = x$. What differs from the BSM case is the mapping of binary sequences, which corresponds for the tent map to a Gray mapping, namely,
\beq
\Mc_{\mathrm{Tent}}(\bv) = \frac{1}{2} + \frac{1}{4} \sum_{l=1}^{\infty} \left(- \frac{1}{2} \right)^{l-1} \prod_{m=1}^{l} \left(2 b_m -1 \right).
\eeq

\item {\bf Logistic map}: The logistic map is defined by
\beq
z_{n+1} = 4 z_n(1-z_n).
\eeq
The invariant pdf is the arcsine distribution, given by
\beq
f_{\mathrm{Log}}(x) = \frac{1}{\pi\sqrt{ z \left(1-z \right)}},
\eeq
which corresponds to 
\beq \label{eq:log_nonlinear_transf}
F^{-1}_{\mathrm{Log}}(x) = \cos^2\left(\frac{\pi}{2} \left( 1-x  \right) \right).
\eeq
Since the logistic map can be seen as a nonlinear transformation of the tent map (with conjugation function $F^{-1}_{\mathrm{Log}}(x)$), binary sequences are mapped to chaotic samples according to $\Mc_{\mathrm{Log}}(\bv) = F^{-1}_{\mathrm{Log}} \left(\Mc_{\mathrm{Tent}}(\bv) \right)$.

\end{itemize}

\begin{remark}
It is possible, for certain cases, to describe the proposed scheme without reference to chaos-theoretic terms. For example, when using the BSM map (see \eqref{BSM}), adaptive-size CCM is equivalent to variable-size ASK with natural mapping. However, casting the technique into an application of chaotic communications allows generalizing the scheme at once to an entire family of schemes whose properties can be studied parametrically according to the map features. It is in this way that the CCM based on the logistic map, which overcomes (as we will see) the one based on BSM in terms of performance, could be derived. Thus, the benefits of introducing concepts of chaos theory in the description of adaptive-size CCM consist in the availability of a set of options and, to a certain extent, of a toolbox ready to use for the analysis and design.    
\end{remark}

\subsection{Anytime reliability of adaptive-size CCM \label{sec:anytime}}

In this section, we will always make the hypothesis that there is no feedback from the receiver, so that no bit gets out of the encoder queue, i.e., $\epsilon_n = 1$, for every $n$. As a consequence, $q_n = n$ and no steps of the chaotic map are ever performed. The transmitted symbol sequence depends then on $f$ only through the mapping $\Mc_f$ and the quantizer $ \Qc_f$ (which in turn depend on $F_f$). 

More precisely, let $s_n \left(\bv\right) $ be the  symbol transmitted at time $n$, where we have explicitly denoted the dependence on the input bit sequence. As $\epsilon_n = 1$, we have that $z_n = z_1 = \Mc_f(\bv)$ and $z_{n}^Q = \Qc_f^{(n)}(z_{1})$. Notice that  $\Qc_f^{(n)} = F_f^{-1} \circ \Qc_U^{(n)} \circ F_f$, where $\Qc_U^{(n)}$ is a uniform quantizer on $[0,1]$ with  $2^n$ levels, defined as
\beq
\Qc_U^{(n)} (x) = \frac{2\iota-1}{2^{n + 1 }} \leftrightarrow x \in \left[\frac{\iota-1}{2^{n}} , \frac{\iota}{2^{n}}\right)
\eeq
for $\iota = 1, \dots, 2^n$.
Now
\ifonecol
\beq
z_n^Q = \Qc_f^{(n)}(z_{1}) = F_f^{-1} \left(\Qc_U^{(n)}\left( F_f \left(z_1 \right)\right)\right) = F_f^{-1} \left(\frac{2 \iota_n(\bv)-1}{2^{n+1}} \right),
\eeq 
\else
\beqa
& z_n^Q = \Qc_f^{(n)}(z_{1}) = F_f^{-1} \left(\Qc_U^{(n)}\left( F_f \left(z_1 \right)\right)\right) \nonumber \\
& = F_f^{-1} \left(\frac{2 \iota_n(\bv)-1}{2^{n+1}} \right),
\eeqa
\fi
where $\iota_n(\bv) \in \{1,\dots, 2^n\}$ is the index such that $F_f \left(z_1 \right) \in \left[\frac{\iota_n(\bv)-1}{2^{n}} , \frac{\iota_n(\bv)}{2^{n}}\right)$. Thus, the transmitted symbol at time $n$ will be given by
\beq
s_n(\bv) = \Gamma_n \left(  F_f^{-1}\left(  \frac{2 \iota_n(\bv)-1}{2^{n+1}}\right) - m_n \right),
\eeq
where, for brevity, $m_n = \EE\{z_n^Q\}$.   

Consider now the $n$-th bit in the information sequence, $b_n$, and let $\widehat{b}_n^{n+d-1} $ be the receiver estimate of $b_n$ at time $n+d-1$. We define $P_n^d(e) \triangleq \PP\{ \widehat{b}_n^{n+d-1} \neq b_n \} $. In the hypothesis that there is no feedback from the receiver, 
the probability $P_n^d(e)$ can be rewritten as
\beq
P_n^d(e) = \frac{1}{2^{n+d-1}} \sum_{\bv_1^{n+d-1}} \PP\{ \widehat{b}_n^{n+d-1} \neq b_n | \bv_1^{n+d-1} \}.
\eeq

In the following, we will derive an upper bound on $P_n^d(e)$ based on the tangential-sphere bound (TSB). Let $\sv(\bv_1^{n+d-1})$ be the sequence of symbols transmitted up to time $n+d-1$. Also, let $\rv_1^{n+d-1}$ be the corresponding vector of received samples. The idea is to upper-bound the probability of incorrectly decoding bit $b_n$ by computing the probability that the received vector is outside a sphere of radius $\rho(n,d,\bv_1^{n+d-1})$ centered on the transmitted symbol sequence $\sv(\bv_1^{n+d-1})$, i.e.,
\ifonecol
\beq \label{eq:TSB}
P_n^d(e) \leq   \frac{1}{2^{n+d-1}} \sum_{\bv_1^{n+d-1}}  \PP \left\{\|\rv_1^{n+d-1} - \sv(\bv_1^{n+d-1})\|_2^2 \geq \rho(n,d,\bv_1^{n+d-1})^2 \right\} ,
\eeq
\else
\beqa \label{eq:TSB}
& P_n^d(e) \leq   \frac{1}{2^{n+d-1}} \sum_{\bv_1^{n+d-1}}  \PP \left\{\|\rv_1^{n+d-1} - \sv(\bv_1^{n+d-1})\|_2^2 \right. \nonumber \\
& \geq \left. \rho(n,d,\bv_1^{n+d-1})^2 \right\} ,
\eeqa
\fi
where $\rho(n,d,\bv_1^{n+d-1})$ is chosen so that all points within the sphere centered on $\sv(\bv_1^{n+d-1})$ with that radius lead to correct decoding of bit $b_n$. 
The following proposition specifies a possible value for $\rho(n,d,\bv_1^{n+d-1})$.

\ifonecol
\begin{lemma} \label{prop:radius}
For an adaptive-size CCM scheme, the TSB \eqref{eq:TSB} holds true when choosing 
\beq \label{eq:rho_new}
\rho(n,d,\bv_1^{n+d-1}) = \rho(n,d,\bv_1^{n}) = \left\{
\begin{array}{ll}
\bar{\rho}\left(n,d, 1 \right), & \iota_n(\bv_1^{n}) = 1, \\
\min\{\bar{\rho}\left(n,d, \iota_n(\bv_1^{n}) \right), \bar{\rho}\left(n,d, \iota_n(\bv_1^{n})-1 \right)\}, & 1 < \iota_n(\bv_1^{n}) < 2^n, \\
\bar{\rho}\left(n,d, 2^n-1 \right), & \iota_n(\bv_1^{n}) = 2^n.
\end{array} \right.
\eeq
where we have defined, for $\iota = 1,\dots, 2^n-1$,
\beq \label{eq:rho_bar}
\bar{\rho}\left(n,d, \iota \right) = \frac1{2} \sqrt{\sum_{j=n}^{n+d-1} \Gamma_j^2 \left( F_f^{-1}\left( \frac{\iota}{2^n} + \frac1{2^{j+1}}\right) - F_f^{-1}\left( \frac{\iota}{2^n} - \frac1{2^{j+1}}\right) \right)^2}.
\eeq
\end{lemma}
\else
\newcounter{MYtempeqncnt}
\begin{figure*}[!t]
\beq \label{eq:rho_new}
\rho(n,d,\bv_1^{n+d-1}) = \rho(n,d,\bv_1^{n}) = \left\{
\begin{array}{ll}
\bar{\rho}\left(n,d, 1 \right), & \iota_n(\bv_1^{n}) = 1, \\
\min\{\bar{\rho}\left(n,d, \iota_n(\bv_1^{n}) \right), \bar{\rho}\left(n,d, \iota_n(\bv_1^{n})-1 \right)\}, & 1 < \iota_n(\bv_1^{n}) < 2^n, \\
\bar{\rho}\left(n,d, 2^n-1 \right), & \iota_n(\bv_1^{n}) = 2^n.
\end{array} \right.
\eeq
\hrulefill
\end{figure*}
\begin{figure*}[!t]
\beq \label{eq:rho_bar}
\bar{\rho}\left(n,d, \iota \right) = \frac1{2} \sqrt{\sum_{j=n}^{n+d-1} \Gamma_j^2 \left( F_f^{-1}\left( \frac{\iota}{2^n} + \frac1{2^{j+1}}\right) - F_f^{-1}\left( \frac{\iota}{2^n} - \frac1{2^{j+1}}\right) \right)^2}.
\eeq
\hrulefill
\end{figure*}
\begin{lemma} \label{prop:radius}
For an adaptive-size CCM scheme, the TSB \eqref{eq:TSB} holds true when choosing $\rho(n,d,\bv_1^{n+d-1})$ as \eqref{eq:rho_new} (shown at the top of next page), where we have defined, for $\iota = 1,\dots, 2^n-1$, $\bar{\rho}\left(n,d, \iota \right)$ as \eqref{eq:rho_bar} (also at the top of next page).
\end{lemma}
\fi

\pf {\ifarxiv
The proof is reported in Appendix~\ref{sec:app1}.
\else
The proof is reported in \cite{ThisPaperArxiv}.
\fi
\qedsymb

Notice that, whenever $F_f(x) = x$, $\forall x \in [0,1]$, the expression of $\rho(n,d,\bv_1^{n})$ simplifies to
\beq \label{eq:rho_for_BSM}
\rho(n,d,\bv_1^{n}) = \bar{\rho}(n,d,1) = \bar{\rho}(n,d) = \frac1{2} \sqrt{\sum_{j=n}^{n+d-1} \Gamma_j^2 4^{-j}}.
\eeq

By substituting the value of the radius specified in Lemma \ref{prop:radius} into \eqref{eq:TSB}, the obtained TSB on $P_n^d(e)$ reads
\beq \label{eq:tsb}
P_n^d(e) \leq  \frac{1}{2^{n}} \sum_{\bv_1^{n}}   \frac{\Gamma\left(\frac{d}{2}, \frac{\rho(n,d,\bv_1^{n})^2}{2 \sigma^2}\right)}{\Gamma\left(\frac{d}{2}\right)},
\eeq
where $\Gamma(k,x)$ is the upper incomplete gamma function. The following proposition provides a sufficient condition for anytime reliability of the adaptive-size CCM.

\begin{proposition} \label{prop:anytime}
Consider an adaptive-size CCM scheme with the following properties.
\begin{itemize}
\item The invariant pdf $F_f^{-1}(x)$ satisfies 
\beq \label{eq:deriv_cond}
\inf_{x \in (0,1)} \frac{\dd F_f^{-1}(x)}{\dd x} > 0 .
\eeq

\item The normalization constant $\Gamma_n$ satisfies $\Gamma_n \geq \lambda 2^n$, for $\lambda > 0$.

\end{itemize}
Then, the following facts holds. 
\begin{enumerate}[i)]
\item For every $n$ and $d \geq d_0 > 2$, $\min_{\bv_1^{n}} \rho(n,d,\bv_1^{n})^2 \geq \beta d $, with $\beta > 0$. 

\item For $\sigma^2 < \sigma^2_{\sup}$, the adaptive-size CCM is anytime reliable with anytime exponent satisfying
\beq 
\gamma \geq \overline{\gamma} = \frac1{2} \left( \frac{\beta}{\sigma^2} - \log \frac{2 \beta d_0 \ee}{(d_0-2) \sigma^2}\right) > 0,
\eeq
where $\sigma^2_{\sup}$ is the unique solution smaller than $\beta$ of the following equation:
\beq \label{eq:new_cond}
\frac{\beta}{\sigma^2_{\sup}} - \log \frac{\beta}{\sigma^2_{\sup}} = \log \frac{2 d_0 e}{d_0-2}.
\eeq

\end{enumerate}
\end{proposition}

\pf i) From \eqref{eq:rho_bar}, in the hypotheses of the proposition, we can bound $\left(\bar{\rho}\left(n,d, \iota \right)\right)^2$ as follows.
\ifonecol
\begin{eqnarray}
\left(\bar{\rho}\left(n,d, \iota \right)\right)^2 & = & \frac1{4} \sum_{j=n}^{n+d-1} \Gamma_j^2 \left( F_f^{-1}\left( \frac{\iota}{2^n} + \frac1{2^{j+1}}\right) - F_f^{-1}\left( \frac{\iota}{2^n} - \frac1{2^{j+1}}\right) \right)^2 \non
& \geq & d \frac{\lambda^2}{4} \min_{j=n}^{n+d-1} \frac{\left( F_f^{-1}\left( \frac{\iota}{2^n} + \frac1{2^{j+1}}\right) - F_f^{-1}\left( \frac{\iota}{2^n} - \frac1{2^{j+1}}\right) \right)^2}{4^{-j}} \non
& \geq & d \frac{\lambda^2}{4} \min_{j=n}^{\infty} \frac{\left( F_f^{-1}\left( \frac{\iota}{2^n} + \frac1{2^{j+1}}\right) - F_f^{-1}\left( \frac{\iota}{2^n} - \frac1{2^{j+1}}\right) \right)^2}{4^{-j}} \triangleq \beta_{n,\iota} d ,
\end{eqnarray}
\else
\begin{eqnarray}
&\left(\bar{\rho}\left(n,d, \iota \right)\right)^2 =  \nonumber \\
& \frac1{4} \sum_{j=n}^{n+d-1} \Gamma_j^2 \left( F_f^{-1}\left( \frac{\iota}{2^n} + \frac1{2^{j+1}}\right) - F_f^{-1}\left( \frac{\iota}{2^n} - \frac1{2^{j+1}}\right) \right)^2 \nonumber \\
& \geq  d \frac{\lambda^2}{4} \min_{j=n}^{n+d-1} \frac{\left( F_f^{-1}\left( \frac{\iota}{2^n} + \frac1{2^{j+1}}\right) - F_f^{-1}\left( \frac{\iota}{2^n} - \frac1{2^{j+1}}\right) \right)^2}{4^{-j}} \nonumber \\
& \geq  d \frac{\lambda^2}{4} \min_{j=n}^{\infty} \frac{\left( F_f^{-1}\left( \frac{\iota}{2^n} + \frac1{2^{j+1}}\right) - F_f^{-1}\left( \frac{\iota}{2^n} - \frac1{2^{j+1}}\right) \right)^2}{4^{-j}} \nonumber \\
& \triangleq \beta_{n,\iota} d ,
\end{eqnarray}
\fi
where the fraction above is the squared difference quotient of function $F_f^{-1}(x)$ computed in $x = \iota 2^{-n} - 2^{-j-1}$. Let $\beta = \min_{n,\iota}   \beta_{n,\iota}$. Then $\beta > 0$ since $F_f^{-1}(x)$ is increasing and its derivative is bounded away from zero in every point of its domain, thanks to \eqref{eq:deriv_cond}. Taking $\rho(n,d,\bv_1^{n})$ as in Lemma \ref{prop:radius} yields the proof of i).

ii) It is easy to verify that, if \eqref{eq:new_cond} is satisfied, then $\beta/\sigma^2_{\sup} > \log 2 $. Thus, we have, for $\sigma^2 < \sigma^2_{\sup}$ and for every $\bv_1^n$
\beq 
\frac{\rho(n,d,\bv_1^{n})^2}{2 \sigma^2} \geq  \frac{\beta d}{2 \sigma^2} > \frac{\beta d}{2 \sigma^2_{\sup}} > \frac{d}{2} \log 2.
\eeq

As a consequence, we can use the following upper bound that can be  found in  \cite{Jameson16}, which holds for $x > a \log 2$,
\beq 
\Gamma (a,x) \leq 2^a x^{a-1} e^{-x}
\eeq
to upper-bound the numerator of each term in \eqref{eq:tsb} as follows:
\ifonecol
\beq 
\Gamma\left(\frac{d}{2}, \frac{\rho(n,d,\bv_1^{n})^2}{2 \sigma^2}\right) \leq \Gamma\left(\frac{d}{2}, \frac{\beta d}{2 \sigma^2}\right) \leq 2^{d/2} \left(\frac{\beta d}{2 \sigma^2} \right)^{d/2-1} \ee^{-\frac{\beta d}{2 \sigma^2}}.
\eeq
\else
\beqa
& \Gamma\left(\frac{d}{2}, \frac{\rho(n,d,\bv_1^{n})^2}{2 \sigma^2}\right) \leq \Gamma\left(\frac{d}{2}, \frac{\beta d}{2 \sigma^2}\right) \nonumber \\
& \leq 2^{d/2} \left(\frac{\beta d}{2 \sigma^2} \right)^{d/2-1} \ee^{-\frac{\beta d}{2 \sigma^2}}.
\eeqa
\fi
Moreover, the following  lower bound to the gamma function $\Gamma(a)$ for $a \geq 1$ can also be derived from \cite{Jameson16}
\beq 
\Gamma\left(a\right) \geq \Gamma\left(a, a-1\right) \geq (a-1)^{a-1}  \ee^{-(a-1)}.
\eeq
Substituting both the upper and the lower bound into \eqref{eq:tsb}, we obtain the following looser upper bound to $P_n^d(e)$
\ifonecol
\begin{eqnarray}
P_n^d(e) \leq & 2\left(  \frac{2 \beta d \ee}{(d-2)\sigma^2}\right)^{d/2-1} \ee^{-\frac{\beta d}{2 \sigma^2}} 
\leq  2\left(  \frac{2 \beta d_0 \ee}{(d_0-2)\sigma^2}\right)^{d/2-1} \ee^{-\frac{\beta d}{2 \sigma^2}} 
=  \frac{(d_0-2)\sigma^2}{\beta d_0 \ee} \ee^{-\overline{\gamma} d},
\end{eqnarray}
\else
\begin{eqnarray}
& P_n^d(e) \leq  2\left(  \frac{2 \beta d \ee}{(d-2)\sigma^2}\right)^{d/2-1} \ee^{-\frac{\beta d}{2 \sigma^2}} \non
& \leq  2\left(  \frac{2 \beta d_0 \ee}{(d_0-2)\sigma^2}\right)^{d/2-1} \ee^{-\frac{\beta d}{2 \sigma^2}} =  \frac{(d_0-2)\sigma^2}{\beta d_0 \ee} \ee^{-\overline{\gamma} d},
\end{eqnarray}
\fi
which implies that the scheme is anytime reliable with an anytime exponent not lower than $\overline{\gamma}$. The fact that $\overline{\gamma}$ is larger than zero follows from \eqref{eq:new_cond} and from the fact that $x - \log x$ is monotonically increasing for $x>1$.
\qedsymb

It is worth noting that the logistic map does not satisfy the conditions of Proposition \ref{prop:anytime}, since
\beq
\frac{\dd F_f^{-1}(x)}{\dd x} = \frac{\pi}{2} \sin \left(\pi x\right),
\eeq
which has a vanishing derivative for $x \rightarrow 0, 1$. 


\subsection{Average energy expenditure}

The condition $\Gamma_n \geq \lambda 2^n$ of Proposition \ref{prop:anytime} implies that the instantaneous power increases exponentially with $n$ when there is no feedback from the receiver. However, when feedback is allowed, the instantaneous power depends on $q_n$, which can be kept small provided that the noise variance is low enough.

In this paragraph, we analyze the average energy expenditure of anytime reliable adaptive-size CCM. More precisely, we will consider the worst-case scenario in which, for every $n$ and $d \geq d_0$,
\beq \label{eq:error_prob_AR_worst}
P_n^d(e) =  K \ee^{-\gamma d},
\eeq
and derive bounds on the distribution of $q_n$, which is the modulation efficiency at time $n$, when feedback from the receiver allows to shorten the input bit queue. The result is given in the following proposition.

\begin{proposition}
For an anytime reliable adaptive-size CCM scheme satisfying \eqref{eq:error_prob_AR_worst}, when there is feedback from the receiver, the modulation efficiency distribution at time $n+d-1$, $n  = 1, 2, \dots$, satisfies
\beq \label{eq:UB_qn_dist}
\PP\{ q_{n+d-1} = d \} \leq K \left( 1+ \frac{\ee^{-2\gamma}}{1 - \ee^{-\gamma}}\right) \ee^{- \gamma d},
\eeq 
for $d \geq d_0$.
\end{proposition}
\pf We first find an upper and a lower bound to $\PP\{ q_{n+d-1} \geq d \}$ for $d \geq d_0$ and then we use such bounds to derive \eqref{eq:UB_qn_dist}.

We have, for $d \geq d_0$,
\ifonecol
\begin{eqnarray} \label{eq:UB_qn_cdf}
\PP\{ q_{n+d-1} \geq d \} & = & \PP\left \{ \bigcup_{j=1}^{n} \{ \widehat{b}_j^{n+d-1}  \neq b_j \} \right\} 
 \leq  \sum_{j=1}^n P_j^{n+d-j}(e) = \sum_{j=1}^n K \ee^{-\gamma (d+j-1)} \non
& \leq & \sum_{j=1}^{\infty} K \ee^{-\gamma (d+j-1)} = \frac{K}{1 - \ee^{-\gamma}} \ee^{- \gamma d},
\end{eqnarray}
\else
\begin{eqnarray} \label{eq:UB_qn_cdf}
& \PP\{ q_{n+d-1} \geq d \}  = \PP\left \{ \bigcup_{j=1}^{n} \{ \widehat{b}_j^{n+d-1}  \neq b_j \} \right\} \non
& \leq  \sum_{j=1}^n P_j^{n+d-j}(e) = \sum_{j=1}^n K \ee^{-\gamma (d+j-1)} \non
& \leq \sum_{j=1}^{\infty} K \ee^{-\gamma (d+j-1)} = \frac{K}{1 - \ee^{-\gamma}} \ee^{- \gamma d},
\end{eqnarray}
\fi
where the first inequality is the union bound. Moreover
\beq \label{eq:LB_qn_cdf}
\PP\{ q_{n+d-1} \geq d \} \geq P_n^{n+d-j}(e) = K \ee^{-\gamma d}.
\eeq

Using \eqref{eq:UB_qn_cdf} and \eqref{eq:LB_qn_cdf}, we obtain, for $d \geq d_0$
\ifonecol
\begin{eqnarray} \label{eq:UB_qn_pdf}
\PP\{ q_{n+d-1} = d \} & = & \PP\{ q_{n+d-1} \geq d \} - \PP\{ q_{n+d-1} \geq d+1 \} 
\leq  K \left( 1+ \frac{\ee^{-2\gamma}}{1 - \ee^{-\gamma}}\right) \ee^{- \gamma d}.
\end{eqnarray}
\else
\begin{eqnarray} \label{eq:UB_qn_pdf}
& \PP\{ q_{n+d-1} = d \}  =  \PP\{ q_{n+d-1} \geq d \} \non
& - \PP\{ q_{n+d-1} \geq d+1 \}
\leq  K \left( 1+ \frac{\ee^{-2\gamma}}{1 - \ee^{-\gamma}}\right) \ee^{- \gamma d}.
\end{eqnarray}
\fi
\qedsymb

With the above result, we can easily bound the average energy expenditure at time $n+d-1$. Notice that, because of ergodicity, such result is useful also to bound the mean energy expenditure over time for a given realization of the adaptive-size CCM scheme. To make things simple, we will suppose that $\Gamma_{d} = \Gamma_0 2^d$, which satisfies the conditions of Proposition \ref{prop:anytime}. Let $E_s(d)$ be the symbol energy for a modulation efficiency of $d$. Since all symbols have a magnitude smaller than $\Gamma_0 2^d$, we have that 
$E_s(d) \leq \Gamma_0^2 4^d$.
Thus, we have
\beq
E_s = \EE_d\left[E_s\left(d\right)\right] \leq E_0 + K \Gamma_0^2 \left( 1+ \frac{\ee^{-2\gamma}}{1 -  \ee^{-\gamma}}\right) \sum_{d=d_0}^{\infty} 4^d \ee^{- \gamma d},
\eeq
where $E_0$ is the average energy for $d < d_0$ and is clearly finite. The second term is finite as long as $\gamma > \ln 4$, in which case
\beq
E_s  \leq E_0 + K  \Gamma_0^2\left( 1+ \frac{\ee^{-2\gamma}}{1 - \ee^{-\gamma}}\right) \frac{(4\ee^{-\gamma})^{d_0}}{1-4\ee^{-\gamma}}.
\eeq

Analogously, for the $m$-th moment,
\beq
\EE_d\left[E_s\left(d\right)^m \right] \leq \mu_{m,0} + K \Gamma_0^{2m} \left( 1+ \frac{\ee^{-2\gamma}}{1 -  \ee^{-\gamma}}\right) \frac{(4^m \ee^{-\gamma})^{d_0}}{1-4^m \ee^{-\gamma}},
\eeq
as long as $\gamma > m\ln 4$.

\section{adaptive-bandwidth CCM} \label{sec:UBCCM}

In this section, we propose a different time-varying CCM scheme, in which, instead of increasing the modulation size, we use an orthogonal modulation to accommodate all bits transmitted at a given time. Each bit then is allocated to a different channel and, within this channel, it is modulated by one of two possible chaotic sequences. From the practical point of view, such adaptive-bandwidth CCM system could be implemented as an adaptive-size FSK one.

For a given chaotic map $f$, the system works as showed in Algorithm 2. The definition of $q_n$ and $\epsilon_n$ is the same as in Algorithm 1.

\begin{table}
\algo{2}{
Adaptive-bandwidth CCM.
 
\begin{itemize}

\item A pair of initial conditions are set to $z^{(0)} = \Mc_f \left( \uv_0 \right)$, $z^{(1)} = \Mc_f \left( \uv_1 \right)$ for suitable semi-infinite binary sequences $ \uv_0$ and  $\uv_1$. Moreover, $\epsilon_0 = 1$.

\item Then, for $n = 1, 2, \dots$ 

\begin{enumerate}

\item Set $q_n = n - \epsilon_{n-1}+1$. The transmitted symbol $\sv_n$ is a size-$q_n$ vector
\beq
\label{s_vector}
 \sv_n = \left( s_n^1 ,\cdots, s_n^{q_n}\right).
\eeq
where
\beq 
s_n^i = f^{(i)} \left( z^{(b_{n-i+1})}\right) = \left\{
\begin{array}{ll}
f\left( s_{n-1}^{i-1}\right), & i > 1, \\
f \left( z^{(b_{n})}\right), & i = 1 .
\end{array} \right.
\eeq

\item The components of $\sv_n$ are transmitted through orthogonal channels. Each of these components is a sample taken from one of two possible chaotic sequences, and carries information about a single bit.  

\item The receiver performs optimal ML decoding and sends back to the transmitter through the feedback channel the value of $\epsilon_{n}$.

\end{enumerate}

\end{itemize}}
\end{table}

If each orthogonal channel is AWGN, the received sample at time $n$ will be
\beq
 \mathbf{r}_{n}=\left(r_{n}^1,\cdots,r_{n}^{q_n}\right)=\mathbf{s}_{n}+\mathbf{w}_{n},
\eeq
where $\mathbf{w}_{n}=\left(w_{n}^1,\cdots,w_{n}^{q_n}\right)$ is a $q_n$-dimensional zero-mean i.i.d. Gaussian noise vector, with noise power per dimension $\sigma^2$.
In this situation, the ML decoding of each bit will consist in comparing the received trajectory with the two possible chaotic sequences. The decoding rule for  bit $b_n$ after $d$ time steps will thus be
\beq
\widehat{b}_n^{n+d-1} = \arg \min_{b_n \in \{0,1\}} \left\{\sum_{j=1}^{d} \left(r_{n+j-1}^j - f^{(j)}\left(z^{(b_n)}\right) \right)^2 \right\}.
\eeq

\begin{remark}
The idea to use a time-varying multi-dimensional system may be also implemented in the simple form of a binary repetition coding scheme with time-varying efficiency. Under certain circumstances (to be demonstrated for the chaos-based system in the sequel) both kinds of systems would offer growing reliability with a growing number of channel uses, but the usage of a chaotic waveform offers additional advantages with respect to the classical counterpart. As mentioned before, a CCM system offers the availability of a whole family of chaotic systems that may be tailored to meet different statistical properties. For example, there is the possibility of self-synchronisation at the receiver, without the need of additional signals or protocols, due to the properties of chaotic waveforms of low autocorrelation out of the origin. Moreover, they may offer as well low probability of interception in absence of third-party knowledge of the specific chaotic dynamics.
\end{remark}

\subsection{Anytime reliability of adaptive-bandwidth CCM \label{sec:ar_ubccm}}

In this subsection, we study the anytime reliability properties of the adaptive-bandwidth CCM, as described above. 
Due to the intrinsic symmetry, the probability that $\hat{b}_n^{n+d-1}$ is not equal to $ b_n$ can be written as
%
\ifonecol
\beq
P_n^d(e)=\PP\left\{\sum_{j=1}^{d} \left(r_{n+j-1}^1 - f^j\left(z^0\right) \right)^2 > \sum_{j=1}^{d} \left(r_{n+j-1}^1 - f^j\left(z^1\right) \right)^2 | b_n=0 \right\},
\eeq
\else
\beqa
& P_n^d(e)=\PP\left\{\sum_{j=1}^{d} \left(r_{n+j-1}^1 - f^j\left(z^{(0)}\right) \right)^2 \right. \non
& > \left. \sum_{j=1}^{d} \left(r_{n+j-1}^1 - f^j\left(z^{(1)}\right) \right)^2 | b_n=0 \right\},
\eeqa
\fi
which, after some algebra, leads to
\beq
P_n^d(e)=\frac{1}{2}  \mathrm{erfc} \left( \frac{d_E\left(d\right)}{2 \sqrt{2 \sigma^2}} \right),
\eeq
where
\beq 
d_E^2\left(d\right)=\sum_{j=1}^{d} \left( f^j\left(z^{(1)}\right) - f^j\left(z^{(0)}\right) \right)^2 = \sum_{j=1}^{d} d_j^2
\eeq
is the squared Euclidean distance between the chaotic trajectory starting at $z^{(1)}$ and the one starting at $z^{(0)}$ after $d$ steps, and $d_j^2$ represents the individual quadratic difference at instant $j$. Using the bound $\mathrm{erfc}\left(x\right) \leq \mathrm{e}^{-x^2}$,
\beq
\label{bound_v}
P_n^d(e) \leq \frac{1}{2}  \mathrm{e}^{-\frac{d_E^2\left(d\right)}{8\sigma^2}}.
\eeq
As is already known, anytime reliability conditions would be proven if, for certain $d_0 > 0$ and $d \geq d_0$, it can be shown that
$d_E^2\left(d\right) \geq \beta d$,
for $\beta > 0$, so that the anytime exponent is bounded as $\gamma \geq \overline{\gamma}=\frac{\beta}{8\sigma^2}$. It is evident that all depends on the trajectories $f^j\left(z^{(0)}\right)$ and $f^j\left(z^{(1)}\right)$, and their differences $d_j^2$. To prove or disprove previous condition on $d_E^2\left(d\right)$, an analysis of the orbits generated by $z^{(0)}$ and $z^{(1)}$ over $f$ is necessary.
%
%
%
%
%
Based on the concept of symbolic dynamics, we can provide sufficient conditions to guarantee anytime reliability when considering specific kinds of maps.

\begin{proposition} \label{prop:AR_bsm}
For the BSM and conjugates of it through a monotonic and strictly increasing conjugation function $g\left(x\right): \left[0,1\right] \rightarrow \left[0,1\right]$, anytime reliability can be achieved when
\begin{enumerate}
 \item The semi-infinite binary symbolic sequence $\uv_0 = \uv$ representing the initial condition $z^{(0)}=g\left(\Mc_{\mathrm{BSM}}\left(\uv_0\right)\right)$ does not contain runs of consecutive $0$'s or $1$'s of length higher than a given maximum value $m_r$.
 \item The initial condition $z^{(1)}$ is chosen according to a binary symbolic sequence $\uv_1 = \overline{\uv}$ complementary of the one corresponding to $z^{(0)}$, so that $z^{(1)}=g\left(\Mc_{\mathrm{BSM}}\left(\overline{\uv}\right)\right) = g\left(1-\Mc_{\mathrm{BSM}}\left(\uv\right)\right)$.
\end{enumerate}
In this case, for $d \geq 1$, $d_E^2\left(d\right) \geq \beta d$, with
\beq 
 \label{rho_BSM}
 \beta=\left(g\left(\frac{1}{2}-\frac{1}{2^{m_r+2}}\right) - g\left(\frac{1}{2}+\frac{1}{2^{m_r+2}}\right)\right)^2 > 0.
\eeq
The BSM case corresponds to $g\left(x\right)=x$, with
 $\beta=\frac{1}{4^{m_r+1}}$.
\end{proposition}

\pf 
In the presence of a run of up to $m_r$ bits in $\uv$, the minimum individual squared distance between the two orbits is reached at iteration $j$ when the symbolic sequence of $f^j\left(z^{(0)}\right)$ corresponds to
\ifonecol
\beq 
 \uv_{j}^{\infty}=\left(u_{j+1}=0,u_{j+2}=1, \cdots, u_{j+m_r+1}=1, u_{j+m_r+2}=0, \cdots\right), 
\eeq
\else
\beqa
 & \uv_{j}^{\infty}=\left(u_{j+1}=0,u_{j+2}=1, \cdots, u_{j+m_r+1}=1, \right. \non 
 & \left. u_{j+m_r+2}=0, \cdots\right), 
\eeqa
\fi
or the complementary one. The bits after $u_{j+m_r+2}$ can take any value. This is the point where $\Mc_{\mathrm{BSM}}\left(\uv_{j}^{\infty}\right)$ and $\Mc_{\mathrm{BSM}}\left(\overline{\uv}_{j}^{\infty}\right)$ are closest to the separation point $1/2$. Indeed, we can verify that
\ifonecol
\beq
 \Mc_{\mathrm{BSM}}\left(\uv_{j}^{\infty}\right)=\frac{1}{2}-\frac{1}{2^{m_r+1}}+\Theta\left(m_r\right), \, 0 \leq \Theta\left(m_r\right) \leq \frac{1}{2^{m_r+2}},
\eeq
\else
\beqa
 & \Mc_{\mathrm{BSM}}\left(\uv_{j}^{\infty}\right) =\frac{1}{2}-\frac{1}{2^{m_r+1}}+\Theta\left(m_r\right), \non  
 & 0 \leq \Theta\left(m_r\right) \leq \frac{1}{2^{m_r+2}},
\eeqa
\fi
and therefore
\beq
 \frac{1}{2}-\frac{1}{2^{m_r+1}} \leq \Mc_{\mathrm{BSM}}\left(\uv_{j}^{\infty}\right) \leq \frac{1}{2}-\frac{1}{2^{m_r+2}}.
\eeq
Moreover, it is easy to see that
\beq 
 \frac{1}{2}+\frac{1}{2^{m_r+2}} \leq \Mc_{\mathrm{BSM}}\left(\overline{\uv}_{j}^{\infty}\right) \leq \frac{1}{2}+\frac{1}{2^{m_r+1}}.
\eeq
Given that $f^j\left(z^{(0)}\right)=g\left(\Mc_{\mathrm{BSM}}\left(\uv_{j}^{\infty}\right)\right)$ and $f^j\left(z^{(1)}\right)=g\left(\Mc_{\mathrm{BSM}}\left(\overline{\uv}_{j}^{\infty}\right)\right)$, it is straightforward to verify that
\ifonecol
\beq
 \left|f^j\left(z_s^0\right)-f^j\left(z_s^1\right)\right| \geq \left|g\left(\frac{1}{2}-\frac{1}{2^{m_r+2}}\right) - g\left(\frac{1}{2}+\frac{1}{2^{m_r+2}}\right)\right| > 0,
\eeq
\else
\beqa
 & \left|f^j\left(z_s^0\right)-f^j\left(z_s^1\right)\right| \non 
 & \geq \left|g\left(\frac{1}{2}-\frac{1}{2^{m_r+2}}\right) - g\left(\frac{1}{2}+\frac{1}{2^{m_r+2}}\right)\right| > 0,
\eeqa
\fi
since these are the closest points both trajectories could reach simultaneously under the conditions of the proposition, on opposite sides of the separation point $g\left(1/2\right)$.  
\qedsymb


\begin{proposition} \label{prop:AR_tm}
For the tent map and conjugates of it through a monotonic and strictly increasing conjugation function $g\left(x\right): \left[0,1\right] \rightarrow \left[0,1\right]$, anytime reliability can be achieved when
\begin{enumerate}
 \item The semi-infinite binary symbolic sequence $\uv_0=\uv$ representing the initial condition $z^{(0)}=g\left(\Mc_{\mathrm{TM}}\left(\uv\right)\right)$ is any sequence.
 \item The initial condition $z^{(1)}$ is chosen according to a binary symbolic sequence $\uv_1=\overline{\uv}$ complementary of the one corresponding to $z^{(0)}$, so that $z^{(1)}=g\left(\Mc_{\mathrm{TM}}\left(\overline{\uv}\right)\right)$.
\end{enumerate}
In this case, for $d \geq 1$, $d_E^2\left(d\right) \geq \beta d$, with
\beq
 \label{rho_TM}
 \beta=\underset{x\in\left[\frac{1}{6},\frac{1}{2}\right)}{\inf}\left\{\left(g\left(x\right)-g\left(x+\frac{1}{3}\right)\right)^2\right\} > 0.
\eeq
For the tent map $g\left(x\right)=x$, and in this case
 $\beta=\frac{1}{9}$.
\end{proposition}

\pf 
Considering the mapping corresponding to the tent map, it is easy to verify that, at a given iteration $j$, the difference between mappings would be
\ifonecol
\begin{eqnarray}
 \label{tm_diff}
 &\left|\Mc_{\mathrm{TM}}\left(\uv_{j}^{\infty}\right) - \Mc_{\mathrm{TM}}\left(\overline{\uv}_{j}^{\infty}\right) \right| =\left| \frac{1}{2} \underset{\mathrm{odd}\, l}{\sum_{l=1}^{\infty}} \left(-\frac{1}{2} \right)^{l-1} \prod_{m=j+1}^{j+l} \left(2 u_m -1 \right) \right| & \nonumber\\
 &=\left| \frac{1}{2} \sum_{k=1}^{\infty} \left(\frac{1}{2} \right)^{2k-2} \prod_{m=j+1}^{j+2k-1} \left(2 u_m -1 \right) \right|, &
\end{eqnarray}
\else
\begin{eqnarray}
 \label{tm_diff}
 &\left|\Mc_{\mathrm{TM}}\left(\uv_{j}^{\infty}\right) - \Mc_{\mathrm{TM}}\left(\overline{\uv}_{j}^{\infty}\right) \right| \non
 & =\left| \frac{1}{2} \underset{\mathrm{odd}\, l}{\sum_{l=1}^{\infty}} \left(-\frac{1}{2} \right)^{l-1} \prod_{m=j+1}^{j+l} \left(2 u_m -1 \right) \right| \non
 &=\left| \frac{1}{2} \sum_{k=1}^{\infty} \left(\frac{1}{2} \right)^{2k-2} \prod_{m=j+1}^{j+2k-1} \left(2 u_m -1 \right) \right|,
\end{eqnarray}
\fi
since
\ifonecol
\begin{eqnarray}
 &&\prod_{m=j+1}^{j+l} \left(2 u_m -1 \right) - \prod_{m=j+1}^{j+l} \left(2 \overline{u}_m -1 \right) = 2 \prod_{m=j+1}^{j+l} \left(2 u_m -1 \right), \, \mathrm{odd}\, l,\\
 &&\prod_{m=j+1}^{j+l} \left(2 u_m -1 \right) - \prod_{m=j+1}^{j+l} \left(2 \overline{u}_m -1 \right) = 0, \, \mathrm{even}\, l.
\end{eqnarray}
\else
\begin{eqnarray}
 &\prod_{m=j+1}^{j+l} \left(2 u_m -1 \right) - \prod_{m=j+1}^{j+l} \left(2 \overline{u}_m -1 \right) \non
 & = 2 \prod_{m=j+1}^{j+l} \left(2 u_m -1 \right), \, \mathrm{odd}\, l, \non
 &\prod_{m=j+1}^{j+l} \left(2 u_m -1 \right) - \prod_{m=j+1}^{j+l} \left(2 \overline{u}_m -1 \right) = 0, \, \mathrm{even}\, l. \nonumber
\end{eqnarray}
\fi
The worst case (minimum value) would correspond to a semi-infinite sequence $\uv_{j}^{\infty}$ where the first term in the summation has a given sign, and the remaining terms the opposite one. There is an infinite number of such sequences. Calculating \eqref{tm_diff} for this case yields
\beq
 \underset{\dv_{j}^{\infty}}{\min}\left\{\left| \frac{1}{2} \sum_{k=1}^{\infty} \left(\frac{1}{2} \right)^{2k-2} \prod_{m=j+1}^{j+2k-1} \left(2 u_m -1 \right) \right|\right\} = \frac{1}{3}.
\eeq
This is the situation, under the conditions of the proposition, where the mappings may come closest to each other, and so would do $f^j\left(z^{(0)}\right)=g\left(\Mc_{\mathrm{TM}}\left(\uv_{j}^{\infty}\right)\right)$ and $f^j\left(z^{(1)}\right)=g\left(\Mc_{\mathrm{TM}}\left(\overline{\uv}_{j}^{\infty}\right)\right)$. Therefore, $\forall$ $j>1$, it is verified that, under the conditions of the proposition,
\ifonecol
\beq
 \left| f^j\left(z^{(0)}\right) - f^j\left(z^{(1)}\right) \right| \geq \underset{x\in\left[\frac{1}{6},\frac{1}{2}\right)}{\inf}\left\{\left|g\left(x\right)-g\left(x+\frac{1}{3}\right)\right|\right\} > 0.
\eeq
\else
\beqa
 &\left| f^j\left(z^{(0)}\right) - f^j\left(z^{(1)}\right) \right| \non
 & \geq \underset{x\in\left[\frac{1}{6},\frac{1}{2}\right)}{\inf}\left\{\left|g\left(x\right)-g\left(x+\frac{1}{3}\right)\right|\right\} > 0.
\eeqa
\fi
\qedsymb

\subsection{Average bandwidth and energy expenditure}

For the adaptive-bandwidth CCM,  the modulation efficiency is related to the number of dimensions in the signal vector, and hence to the required bandwidth and energy. Assuming a required bandwidth per dimension of $\Delta f$, the total bandwidth for efficiency $d$ is given by
 $B\left(d\right)=d\Delta f$.
Suppose that the error probability $P_n^d(e)$ satisfies  \eqref{eq:error_prob_AR_worst}. In such a case, with the help of \eqref{eq:UB_qn_dist}, the average bandwidth can be bounded  as
\ifonecol
\begin{eqnarray}
\overline{B} = \EE_d\left[B\left(d\right)\right] \leq 
B_0 + \Delta f K \left(1 + \frac{\ee^{-2\gamma}}{1+\ee^{-\gamma}}  \right) \frac{\ee^{-\gamma d_0} \left(d_0 - \left(d_0 -1 \right) \ee^{-\gamma} \right)}{\left( 1 - \ee^{-\gamma} \right)^2}
\end{eqnarray}
\else
\begin{eqnarray}
& \overline{B} = \EE_d\left[B\left(d\right)\right] \non 
& \leq B_0 + \Delta f K \left(1 + \frac{\ee^{-2\gamma}}{1+\ee^{-\gamma}}  \right) \frac{\ee^{-\gamma d_0} \left(d_0 - \left(d_0 -1 \right) \ee^{-\gamma} \right)}{\left( 1 - \ee^{-\gamma} \right)^2}
\end{eqnarray}
\fi
where $B_0$ is the average bandwidth for $d < d_0$. 
Regarding the average energy expenditure, since the chaotic maps considered here have finite support $\left[0,1\right]$, the energy of vector \eqref{s_vector} satisfies $\sv_n \sv_n^T\leq q_n $.
Therefore, analogously as for the average bandwidth, the average energy can be bounded by  
\beq 
 E_s \leq E_0 + K \left(1 + \frac{\ee^{-2\gamma}}{1+\ee^{-\gamma}}  \right) \frac{\ee^{-\gamma d_0} \left(d_0 - \left(d_0 -1 \right) \ee^{-\gamma} \right)}{\left( 1 - \ee^{-\gamma} \right)^2},
\eeq
where $E_0$ is the average energy for $d < d_0$.

\section{Performance results} \label{sec:performance}

%


\subsection{Performance results for the adaptive-size CCM scheme}

\begin{figure}[htbp]
\centering
\ifonecol
\subfigure[Bit error probability as a function of the decoding delay.]{\label{Fig1}\includegraphics[width=0.49\textwidth,height=67mm]{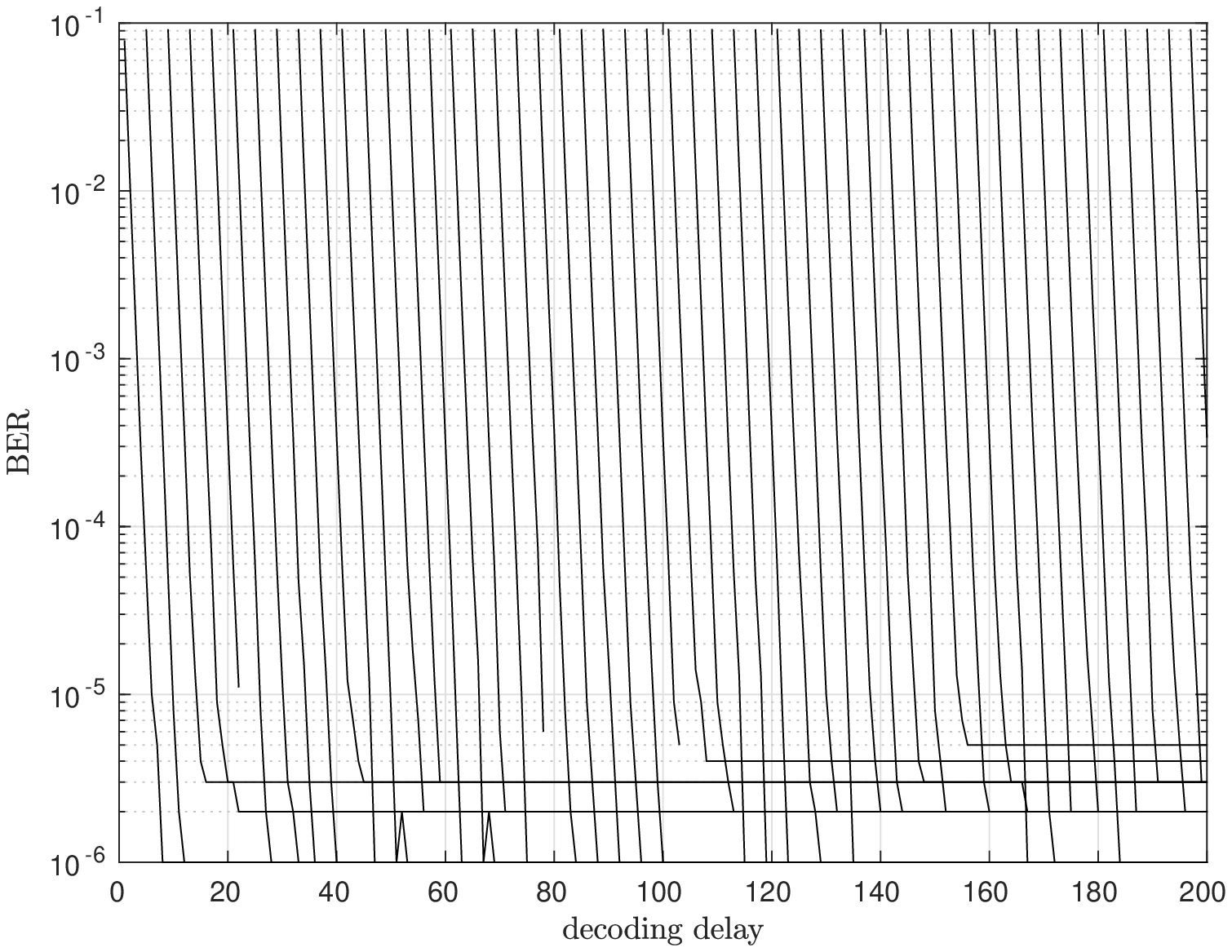}}
\subfigure[Histogram of the modulation efficiency.]{\label{Fig2}\includegraphics[width=0.49\textwidth,height=67mm]{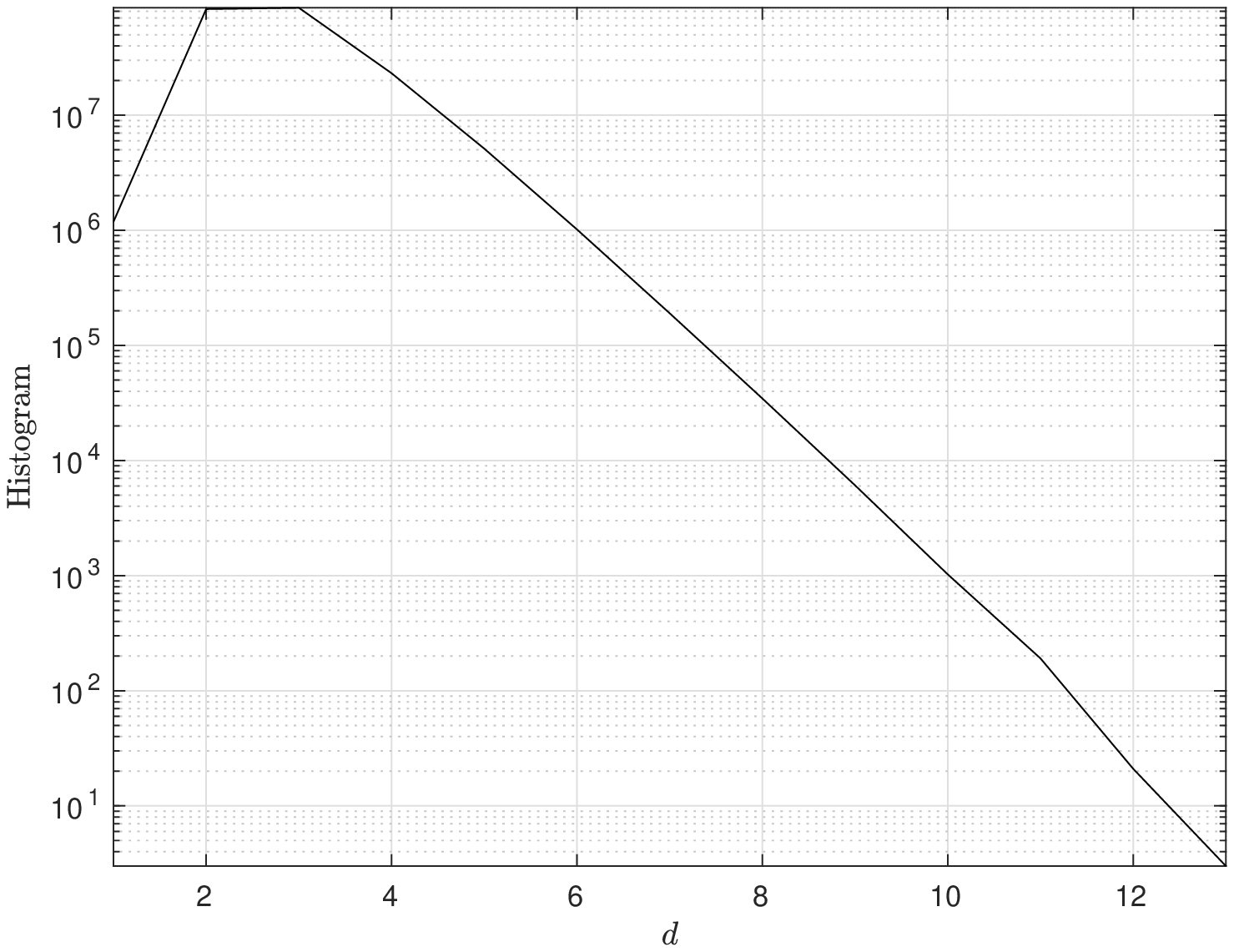}}
\else
\subfigure[Bit error probability as a function of the decoding delay.]{\label{Fig1}\includegraphics[width=\figurewidth]{BER_UP_BSM_N1000000_NP-3dB.eps}}
\subfigure[Histogram of the modulation efficiency.]{\label{Fig2}\includegraphics[width=\figurewidth]{HG_UP_BSM_N1000000_NP-3dB.eps}}
\fi
\caption{BER and efficiency histogram for the adaptive-size CCM scheme.}
\end{figure}
We first consider adaptive-size CCM with the BSM map, when the normalization constant is $\Gamma_j = 2^{j+1}$, $j \geq 1$. In this setting, thanks to \eqref{eq:rho_for_BSM}, the conditions of Prop. \ref{prop:anytime} are satisfied with $\beta = 1$, $d_0 = 3$, $\sigma^2 < \sigma^2_{\sup} \simeq 0.2361$.

\begin{figure}[htbp]
\centering
\ifonecol
\includegraphics[width=0.5\textwidth]{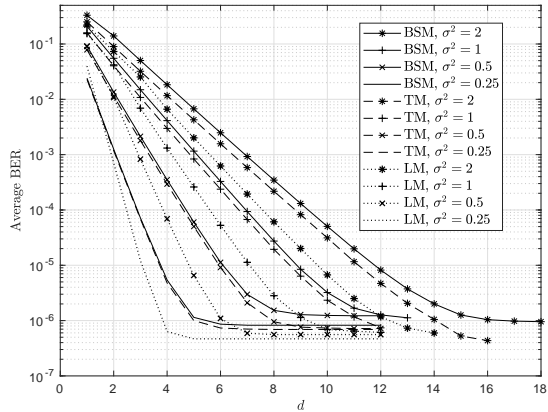}
\else
\includegraphics[width=\figurewidth]{ABER_UP_NPdiff_N1000000.eps}
\fi
\caption{\label{Fig4} Measured average bit error rate for different values of noise power $\sigma^2$ and three different maps: Bernoulli shift map (BSM), tent map (TM) and logistic map (LM).}
\end{figure} 
Figure \ref{Fig1} shows Monte-Carlo results in the case where the noise power is equal to $\sigma^2 = 0.5$ and blocks have a length of $200$ bits. The decoder considers a bit reliably decoded whenever the residual error after decoding (estimated as a function of the magnitude of the output log-likelihood ratio) is equal to $P_e^{\mathrm{res}} = 10^{-5}$. A maximum of $10^6$ blocks have been simulated. Notice that, since for this case $\sigma^2 > \sigma^2_{\sup}$, the conditions of Prop. \ref{prop:anytime} are not satisfied. However, from Figure \ref{Fig1}, which shows the bit error probability as a function of the decoding delay for each bit position in the block, we can see that the scheme is anytime reliable. Indeed, the performance is quite independent on the position in the block, and the bit error probability drops exponentially fast to zero.  We can also see the residual error appearing below $10^{-5}$.

Figure \ref{Fig2} shows the histogram of the modulation efficiency, which, as predicted by the theoretical analysis of Section \ref{sec:VSCCM}, also drops to zero exponentially, for $d \geq 3$. The empirical mean of the modulation efficiency is about $2.76$. In none of the $10^6$ blocks the modulation efficiency went over $13$ bits. 

In Figure \ref{Fig4}, we compare in terms of measured $P_n^{d}(e)$ three different maps, i.e., BSM, tent map and logistic map, for different values of $\sigma^2$ and $10^6$ simulated blocks of length $200$ bits. It can be seen that all curves exhibit an exponential decrease of $P_n^{d}(e)$ with $d$, up to a certain level. Moreover, the slope increases with decreasing values of $\sigma^2$. It is also important to notice that, while BSM and tent map show a similar anytime exponent, the logistic map shows a faster slope (i.e., a larger anytime exponent), which is essentially due to the nonlinear transformation in \eqref{eq:log_nonlinear_transf}. Thus, although the logistic map does not satisfy the conditions of Proposition \ref{prop:anytime}, it is anytime reliable all the same. Finally, notice that we can observe at a certain point an error floor (due to bits incorrectly considered well estimated at the receiver), giving rise to a residual bit error probability, which is however always lower than the target value $P_e^{\mathrm{res}} = 10^{-5}$.    

\subsection{Performance results for the adaptive-bandwidth CCM scheme}

\begin{figure}[htbp]
\centering
\ifonecol
\subfigure[Bit error probability as a function of the decoding delay.]{\label{Fig5}\includegraphics[width=0.49\textwidth,height=67mm]{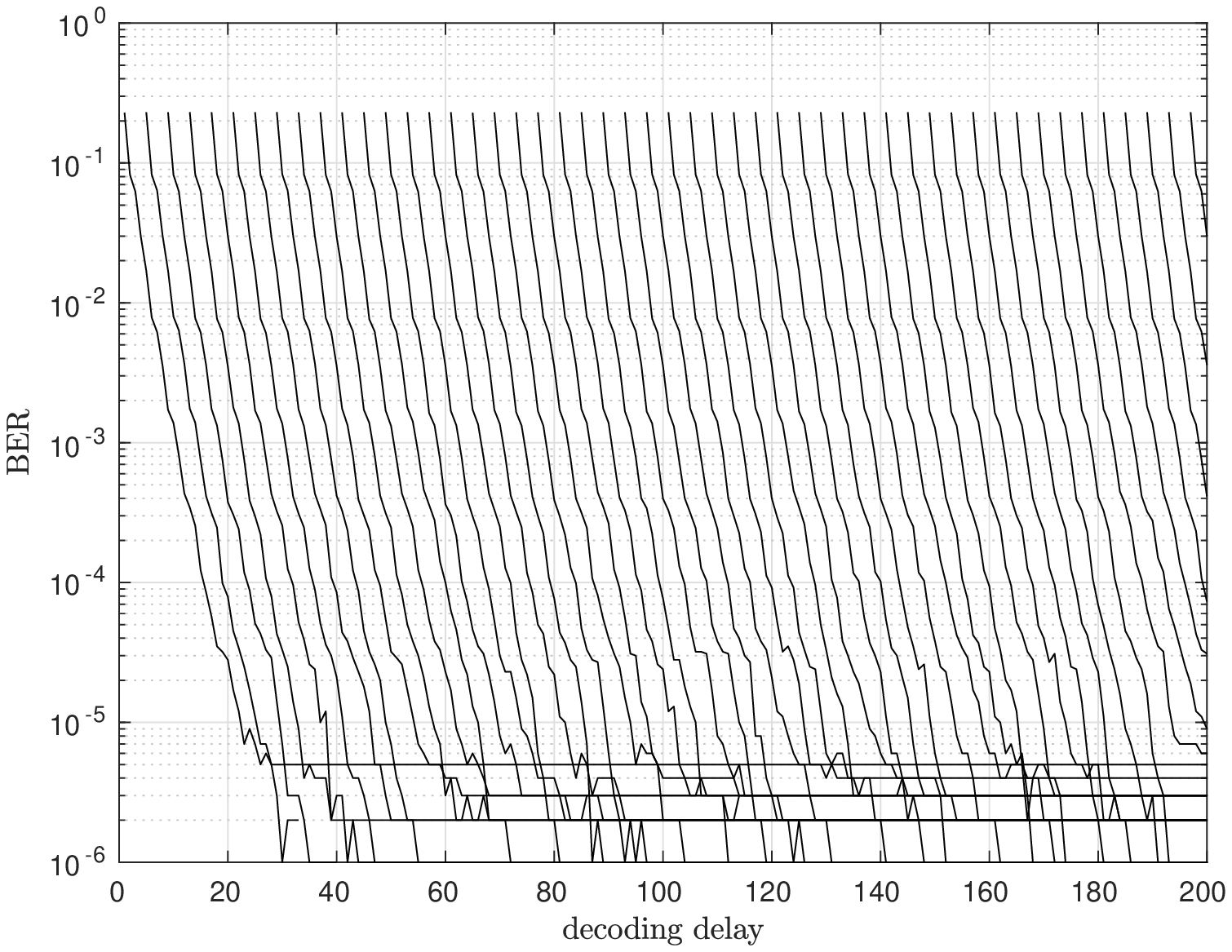}}
\subfigure[Histogram of the modulation efficiency.]{\label{Fig6}\includegraphics[width=0.49\textwidth,height=67mm]{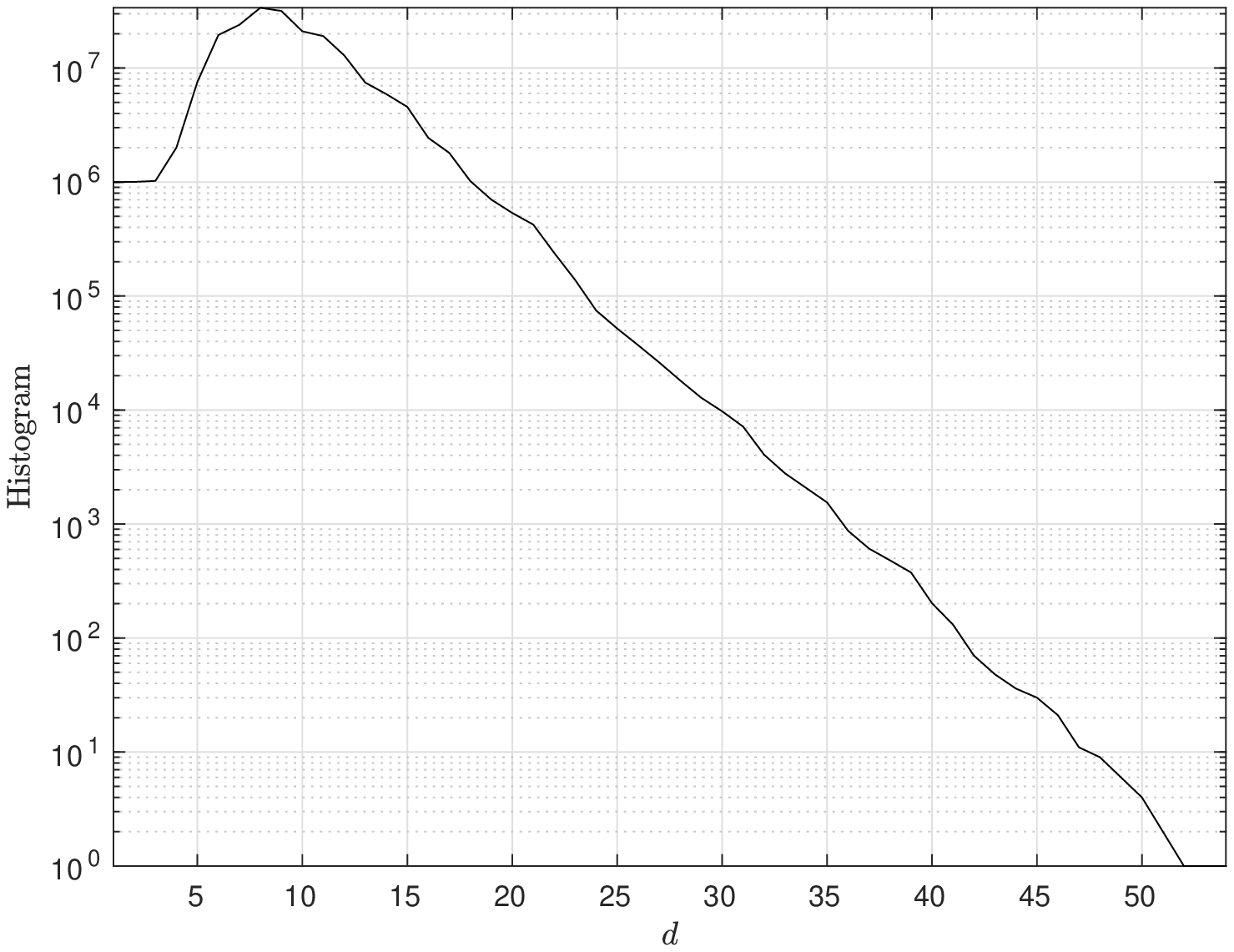}}
\else
\subfigure[Bit error probability as a function of the decoding delay.]{\label{Fig5}\includegraphics[width=\figurewidth]{BER_UB_LOG_N1000000_NP-3dB.eps}}
\subfigure[Histogram of the modulation efficiency.]{\label{Fig6}\includegraphics[width=\figurewidth]{HG_UB_LOG_N1000000_NP-3dB.eps}}
\fi
\caption{BER and efficiency histogram for the adaptive-bandwidth CCM scheme.}
\end{figure}
For exemplification, we choose the logistic map, though the results for the BSM or the tent map would be similar. Using a pair of initial conditions as detailed in Prop. \ref{prop:AR_tm}, we know that anytime reliability conditions are guaranteed with $\gamma\geq \overline{\gamma}=\frac{\beta}{8\sigma^2}$, and $\beta$ as in \eqref{rho_TM}. For the logistic map, $g\left(x\right)=\cos^2\left(\frac{\pi}{2}\left(1-x\right)\right)$ and $\beta \simeq 0.1875$.
%
To make the system practical, we have used random sequences of $N=1000$ bits, and a quantizer with $20$ bits in all the cases. 
Moreover, sequences are normalized to fit in the interval $\left[-1,1 \right]$, so that they are zero-mean. 
 
Figure \ref{Fig5} shows Monte-Carlo results in the case where the noise power is equal to $\sigma^2 = 0.5$, blocks have a length of $200$ bits and the tolerable residual error after decoding is set to $P_e^{\mathrm{res}} = 10^{-5}$. A maximum of $10^6$ blocks have been simulated. From Figure \ref{Fig5}, which shows the bit error probability as a function of the decoding delay for each bit position in the block, we can see that the scheme is anytime reliable, with a residual error below $10^{-5}$. Indeed, the performance is quite independent of the position in the block, and the bit error probability drops exponentially to zero. We can see there are some slope changes for particular bits, specially as we approach and reach the maximum efficiency $d$ attained. This is related to the chaotic nature of the encoding, where the evolution of the log-likelihood ratios (LLRs) at the decoding stage is linked to the evolution of the differences among possible trajectories $\left| f^j\left(z^{(0)}\right) - f^j\left(z^{(1)}\right) \right|$. 

Figure \ref{Fig6} shows the histogram of the modulation efficiency, which, as predicted by the theoretical analysis of Section \ref{sec:UBCCM}, also drops to zero exponentially, now for $d\geq8$. The empirical mean of the modulation efficiency is about $9.25$. As compared to the previous case (adaptive-size CCM with BSM), for the same amount of noise, we attain larger values for the modulation efficiency $d$, and the exponentially decaying trend in the values of the bit error rate is not so steep. This is partly due to the fact that the instantaneous power in the adaptive-size CCM grows exponentially with $d$, whereas for the adaptive-bandwidth CCM it grows linearly. 

\begin{figure}[htbp]
\centering
\ifonecol
\includegraphics[width=0.5\textwidth]{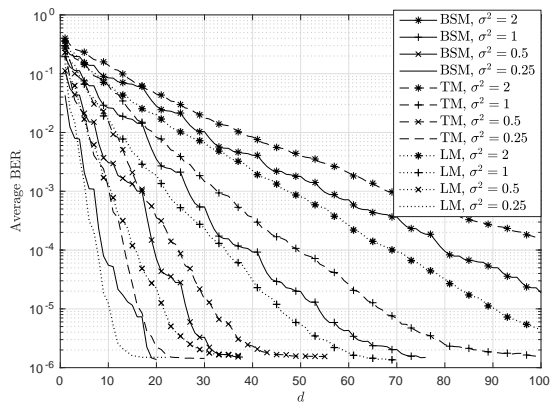}
\else
\includegraphics[width=\figurewidth]{ABER_UB_NPdiff_N1000000.eps}
\fi
\caption{\label{Fig8} Measured average bit error probability for different values of noise power $\sigma^2$ and three different maps: Bernoulli shift map (BSM), tent map (TM) and logistic map (LM).}
\end{figure}
In Figure \ref{Fig8}, we compare in terms of measured $P_n^{d}(e)$ three different maps, i.e., BSM, tent map and logistic map, for different values of $\sigma^2$ and $10^6$ simulated blocks of length $200$ bits. In the case of the BSM, we have chosen a maximal run length of $m_r=5$, so that $\beta\simeq 2.44\cdot 10^{-4}$. It can be seen that, for the particular values of the randomly chosen initial condition pair for each kind of map, the tent map performs poorer than the BSM and the logistic map, which have similar behavior. The best one is the logistic map, for reasons similar to what was seen in the case of the adaptive-size CCM: its nonlinear transform makes it perform better. It is to be noted that $\beta$ is a very conservative parameter: though for the BSM it takes a value several orders of magnitude lower than the one for the tent map ($\beta=1/9$), the experiments show a better trend for the first one. This is related to the fact that its calculation resorts to the worst case possible for the difference among chaotic trajectories, and the frequency of the related event is really small. Notice also that the residual bit error probability is lower than $P_e^{\mathrm{res}}$ for the curves going down to $10^{-5}$. This error floor phaenomenon was already identified in the case of the adaptive-size CCM.

\subsection{Comparison among different CCM schemes}

\ifonecol
\begin{center}
\begin{tabular}{|c|c|c|c|c|c|c||c|c|c|c|c|c|}
\hline
 & \multicolumn{6}{|c||}{adaptive-size CCM} & \multicolumn{6}{|c|}{adaptive-bandwidth CCM}\\
\hline
 & \multicolumn{2}{|c|}{BSM} & \multicolumn{2}{|c|}{TM} & \multicolumn{2}{|c||}{LM} & \multicolumn{2}{|c|}{BSM} & \multicolumn{2}{|c|}{TM} & \multicolumn{2}{|c|}{LM} \\
\hline
 $\sigma^2$ & $\overline{d}$ & std & $\overline{d}$ & std & $\overline{d}$ & std & $\overline{d}$ & std & $\overline{d}$ & std & $\overline{d}$ & std\\
\hline
 $2$ & $4.76$ & $1.27$ & $4.75$ & $1.27$ & $4.28$ & $1.15$ & $53.09$ & $21.30$ & $64.67$ & $28.47$ & $45.49$ & $17.06$\\ 
\hline
 $1$ & $3.80$ & $1.07$ & $3.80$ & $1.08$ & $3.41$ & $0.87$ & $25.30$ & $8.30$ & $32.18$ & $11.39$ & $21.35$ & $7.44$\\
\hline
 $0.5$ & $2.76$ & $0.82$ & $2.76$ & $0.82$ & $2.55$ & $0.67$ & $13.19$ & $4.22$ & $15.75$ & $4.60$ & $9.25$ & $3.09$\\
\hline
 $0.25$ & $2.00$ & $0.52$ & $1.99$ & $0.52$ & $2.01$ & $0.37$ & $5.33$ & $2.14$ & $8.62$ & $2.42$ & $4.52$ & $1.38$\\
\hline
\end{tabular}
\captionof{table}{\label{Table1}Average value of $d$ ($\overline{d}$) and its standard deviation (std) for different cases.}
\end{center}
\else
\begin{center}
\begin{tabular}{|c|c|c|c|c|c|c|}
\hline
 & \multicolumn{6}{|c|}{adaptive-size CCM} \\
\hline
 & \multicolumn{2}{|c|}{BSM} & \multicolumn{2}{|c|}{TM} & \multicolumn{2}{|c|}{LM}  \\
\hline
 $\sigma^2$ & $\overline{d}$ & std & $\overline{d}$ & std & $\overline{d}$ & std \\
\hline
 $1$ & $3.80$ & $1.07$ & $3.80$ & $1.08$ & $3.41$ & $0.87$ \\
\hline
 $0.5$ & $2.76$ & $0.82$ & $2.76$ & $0.82$ & $2.55$ & $0.67$ \\
\hline
 $0.25$ & $2.00$ & $0.52$ & $1.99$ & $0.52$ & $2.01$ & $0.37$ \\
\hline
\hline
 & \multicolumn{6}{|c|}{adaptive-bandwidth CCM}\\
\hline
 & \multicolumn{2}{|c|}{BSM} & \multicolumn{2}{|c|}{TM} & \multicolumn{2}{|c|}{LM}  \\
\hline
 $\sigma^2$ & $\overline{d}$ & std & $\overline{d}$ & std & $\overline{d}$ & std \\
\hline
 $1$ & $25.30$ & $8.30$ & $32.18$ & $11.39$ & $21.35$ & $7.44$\\
\hline
 $0.5$ & $13.19$ & $4.22$ & $15.75$ & $4.60$ & $9.25$ & $3.09$\\
\hline
 $0.25$ & $5.33$ & $2.14$ & $8.62$ & $2.42$ & $4.52$ & $1.38$\\
\hline
\end{tabular}
\captionof{table}{\label{Table1}Average value of $d$ ($\overline{d}$) and its standard deviation (std) for different cases.}
\end{center}
\fi
In Table \ref{Table1}, we can see the mean value of $d$ and its standard deviation for a number of cases. Block length is $200$, for a total number of simulated blocks of $10^4$. Residual error rate for threshold is set to $P_e^{\mathrm{res}}=10^{-5}$.  As we can see, there is a clear advantage in the case of adaptive-size CCM systems, for the same amount of noise power.
 We can also see that the logistic map offers always the best performance, and the rest of trends seen before are confirmed for the given range of $\sigma^2$ values: BSM and tent map perform largely equal in the adaptive-size CCM case, and the first one outperforms the second one in the adaptive-bandwidth CCM case. Note that, for the adaptive-bandwidth system, even a modulation efficiency of several tens does not pose a big technological challenge.

In Table \ref{Table2}, we have the values of the empirical average signal-to-noise ratio (SNR), expressed in dBs, for the two setups considered, and the same kind of maps. The simulation parameters are as before. We can see now the whole picture: while there is an advantage for the adaptive-size CCM from the point of view of the average modulation efficiency, the adaptive-bandwidth CCM case has an advantage in SNR. Its values are always bounded around $10$ dB, whereas for the adaptive-size CCM they explode when the noise power is high. This is related to the fact that the power in this case grows exponentially with $d$, while in the previous case it grows linearly. The trends show that for the adaptive-size CCM a duplication of the noise power is reflected in almost a squaring of the average signal power. In the case of the adaptive-bandwidth CCM, a duplication of the noise power essentially leads to a duplication of the average signal power, and this keeps the average SNR largely constant along the range.
\begin{center}
\begin{tabular}{|c|c|c|c||c|c|c|}
\hline
 & \multicolumn{3}{|c||}{adaptive-size CCM} & \multicolumn{3}{|c|}{adaptive-bw CCM}\\
\hline
 $\sigma^2$ & BSM & TM & LM & BSM & TM & LM \\
\hline
 $1$ & $33.90$ & $32.21$ & $22.93$ & $9.26$ & $10.30$ & $10.28$\\
\hline
 $0.5$ & $19.84$ & $20.08$ & $16.14$ & $9.43$ & $10.20$ & $9.65$\\
\hline
 $0.25$ & $14.52$ & $14.51$ & $13.82$ & $8.49$ & $10.58$ & $9.54$\\
\hline
\end{tabular}
\captionof{table}{\label{Table2}Average SNR (dB) for different cases.}
\end{center}

{\ifarxiv
\else
\color{red}
\fi
It is worth noting that, typically, the pure coding schemes described, e.g.,  in \cite{Grosjean16, Noor-A-Rahim18} have problems whenever the amount of information to be transmitted at each time is very low. For example, Figure 7 of \cite{Noor-A-Rahim18} shows that, when transmitting 16 coded bits per time step, the performance curves of $P_n^d(e)$ show a very relevant slope loss. As an example, for the best performing design, $P_n^d(e) \simeq 7 \times 10^{-5}$ for $d = 50$ at a signal-to-noise ratio of 0.5 dB. Moreover, Figure 14 of the same paper shows that feedback does not dramatically improve the performance on the AWGN channel. On the contrary, our scheme does not suffer from the short length of the information word at each time step. Indeed, our discussion has focused on the particular case where a single information bit is generated at each time step. Such scenario can arise when scalar measurements are differentially encoded, and the transmitted bit is thus simply a flag that denotes whether the state variable is increasing or decreasing. So, in our opinion, our scheme is particularly well suited when the system throughput is very low. On the contrary, pure coding schemes are well suited whenever at each time step there is a considerable information to be conveyed through the channel.}

\ifarxiv
\section{Conclusions}
\else
\section{{\color{red}Conclusions}}
\fi
\label{sec:conclusions}

{\ifarxiv
\else
\color{red}
\fi
In this paper, we have presented two alternatives to define anytime reliable systems using chaos-based communications. In both cases, we have studied the conditions needed for anytime reliability, and we have provided relevant design criteria. We have also developed formal proofs for anytime reliability, under appropriate hypotheses. The study of possible implementations has led to practical trade-offs, and the simulations provided for a variety of cases have shown that the proposed systems work as expected, providing valid approaches to handle communications for anytime reliable systems under AWGN. One of the alternatives is based in driving the instantaneous power of the transmitted sequence, while the other drives the bandwidth required. Both systems offer relative advantages and disadvantages, as illustrated through the simulation results, that also provide useful design hints. In any case, the chaos-based systems considered are simple, and easy to implement. Future work on this respect may address the application of the same principles for other relevant kinds of channels.}


%

\ifarxiv
\appendices
\section{Proof of Lemma \protect\ref{prop:radius} \label{sec:app1}}
Before facing the proof of Lemma \ref{prop:radius}, we prove a simple property of sequences for any adaptive-size CCM scheme. Let  $s_n \left(\bv_1^{n}\right) $ be the  symbol transmitted at time $n$. We remind that there is no feedback from the receiver, so that $\epsilon_n = 1$ for every $n$.
\begin{figure}[htbp]
\centering
\includegraphics[width=0.6\textwidth]{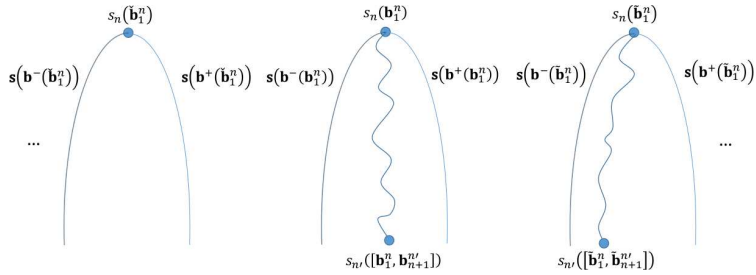}
\caption{\label{FigFO} Forward-ordered adaptive-size CCM scheme.}
\end{figure}

\begin{proposition}[Forward ordering]
Consider a CCM scheme as defined in the previous section. Given $n$, let $\bv_1^{n}$ and $\widetilde{\bv}_1^{n}$ be two different input bit sequences with $s_n\left(\bv_1^{n}\right)  < s_n\left(\widetilde{\bv}_1^{n}\right) $. Then, for all $n' \geq n$ and all binary vectors $\bv_{n+1}^{n'}$ and $\widetilde{\bv}_{n+1}^{n'}$
\beq
s_{n' } \left( \bv_1^{n'} \right) < s_{n' }\left(\widetilde{\bv}_1^{n'}\right).
\eeq 
\end{proposition}

\pf
As $\epsilon_n = 1$, we have that $z_n = z_1 = \Mc_f(\bv)$ and $z_{n}^Q = \Qc_f^{(n)}(z_{1})$. Notice that $\Mc_f(\bv) = F_f^{-1} \left(\Mc_{f_u}(\bv)\right)$, where $f_u$ is the map with uniform invariant density that is topologically conjugate with $f$, and that $\Qc_f^{(n)} = F_f^{-1} \circ \Qc_U^{(n)} \circ F_f$, where $\Qc_U^{(n)}$ is a uniform quantizer on $[0,1]$ with  $2^n$ levels. Thus:
\begin{eqnarray}
z_n^Q &=& F_f^{-1} \left(\Qc_U^{(n)}\left( F_f \left(F_f^{-1} \left(\Mc_{f_u}(\bv)\right)\right)\right)\right) \\
& = & F_f^{-1} \left(\Qc_U^{(n)} \left(z_{1,U}\right)\right)
\end{eqnarray} 
where we have defined $z_{1,U} = \Mc_{f_u}(\bv)$. Since $F_f^{-1}$ is  increasing and the scaling and offset of \eqref{eq:s_n_as} preserves ordering, $s_n\left(\bv_1^{n}\right)  < s_n\left(\widetilde{\bv}_1^{n}\right)$ implies $\Qc_U^{(n)} \left(z_{1,U}\right) < \Qc_U^{(n)} \left(\widetilde{z_{1,U}}\right)$, where $\widetilde{z_{1,U}} = \Mc_{f_u}(\widetilde{\bv})$. For $n'=n$ the proposition is trivial. For $n'>n$, $\Qc_U^{(n')}$ is a finer quantizer than $\Qc_U^{(n)}$. In other words, each of the quantization intervals of $\Qc_U^{(n)}$ is partitioned into $2^{n'-n}$ subintervals, which correspond to the quantization intervals of $\Qc_U^{(n')}$. It trivially follows that $\Qc_U^{(n)} \left(z_{1,U}\right) < \Qc_U^{(n)} \left(\widetilde{z_{1,U}}\right)$ implies $\Qc_U^{(n')} \left(z_{1,U}\right) < \Qc_U^{(n')} \left(\widetilde{z_{1,U}}\right)$ and thus $s_{n' } \left( \bv_1^{n'} \right) < s_{n' }\left(\widetilde{\bv}_1^{n'}\right)$. 
\qedsymb

Notice that the proposition implies immediately that, if $s_n\left(\bv_1^{n}\right) < s_n\left(\widetilde{\bv}_1^{n}\right)$, then for all $n' < n$, $s_{n' }\left(\bv_1^{n'}\right) \leq s_{n' }\left(\widetilde{\bv}_1^{n'}\right)$. A pictorial representation of the forward ordering property is in Figure \ref{FigFO}. As is shown, the ''cone'' of sequences that share the same value of $s_n\left(\bv_1^{n}\right)$ will not cross sequences from other cones. The figure also shows the two extremal symbol sequences of a cone, i.e., those that minimize and maximize the transmitted symbols, which are denoted $\sv \left(\bv^{-}(\bv_1^{n}\right))$ and $\sv \left(\bv^{+}(\bv_1^{n}\right))$ respectively. Formally, they are defined as follows:
\beq
\bv^{+}(\bv_1^{n}) = \left[\bv_1^{n}, \arg \max_{\bv_{n+1}^{n+d-1}} s_{n+d-1}\left( [\bv_1^{n},\bv_{n+1}^{n+d-1}] \right) \right],
\eeq
and
\beq
\bv^{-}(\bv_1^{n}) = \left[\bv_1^{n}, \arg \min_{\bv_{n+1}^{n+d-1}} s_{n+d-1}\left( [\bv_1^{n},\bv_{n+1}^{n+d-1}] \right)\right].
\eeq
For ease of notation, we will omit the dependence on $\bv_1^{n}$ and write simply $\bv^{+}$ and $\bv^{-}$. 

\textit{Proof of Lemma \ref{prop:radius}.} We will prove the lemma by showing that, for each possible input sequence $\bv_1^{n+d-1}$ with $\iota_n(\bv_1^{n}) < 2^n$, all points within the hypersphere centered on $\sv(\bv_1^{n+d-1})$  with radius $\bar{\rho}(n,d,\iota_n(\bv_1^{n}))$ are closer to the symbol sequence $\sv(\bv^+)$ than to any symbol sequence $\sv(\widetilde{\bv})$ with $s_n(\widetilde{\bv}) > s_n(\bv^{+})$. Moreover, for each possible input sequence $\bv_1^{n+d-1}$ with $\iota_n(\bv_1^{n}) > 1$, all points within the hypersphere centered on $\sv(\bv_1^{n+d-1})$  with radius $\bar{\rho}(n,d,\iota_n(\bv_1^{n})-1)$ are closer to the symbol sequence $\sv(\bv^-)$ than to any symbol sequence $\sv(\check{\bv})$ with $s_n(\check{\bv}) < s_n(\bv^{-})$. 
As a consequence, if $\rho(n,d,\bv_1^{n})$ is chosen as in \eqref{eq:rho_new},  the above two conditions are both satisfied and ML decoding will necessarily give $\widehat{b}_n^{n+d-1}  =b_n$ for all points within the hypersphere with radius $\rho(n,d,\bv_1^{n})$.\footnote{For the two extreme cases $\iota_n(\bv_1^{n}) = 1$ and $\iota_n(\bv_1^{n}) = 2^n$, only one of the two conditions must be satisfied, since there is no sequence to the left of $\bv^{-}$ whenever $\iota_n(\bv_1^{n}) = 1$ and there is no sequence to the right of $\bv^{+}$ whenever $\iota_n(\bv_1^{n}) = 2^n$. This is kept into account in \eqref{eq:rho_new}.} 

In order to prove the condition on $\bv^+$, we notice that the received sample vectors at the same distance from $\sv(\bv^+)$ and $\sv(\widetilde{\bv})$ belong to the hyperplane $\pi(\bv^+,\widetilde{\bv})$ given by the equation\footnote{Notice that, since $s_n(\widetilde{\bv}) > s_n(\bv^+)$, $\widetilde{\bv}$ and $\bv^+$ differ in at least one of the first $n$ positions.}
\beq
\pi(\bv^+,\widetilde{\bv}) = \left\{ \rv_1^d: \,\,\, \zeta\left(\rv_1^d, \bv^+, \widetilde{\bv}\right) = 0\right\},
\eeq
with 
\beq
\zeta \left(\rv_1^{n+d-1}, \bv^+, \widetilde{\bv}\right) =  2\sum_{j=1}^{n+d-1} r_j \Gamma_j \Delta_j(\bv^+, \widetilde{\bv})  -  \sum_{j=1}^{n+d-1} \Gamma_j^2  \left( \Sigma_j(\bv^+, \widetilde{\bv})  - 2m_j\right) \Delta_j(\bv^+, \widetilde{\bv}),
\eeq
having defined
\beq
\Delta_j(\bv^+, \widetilde{\bv}) =  F_f^{-1}\left(  \frac{2 \iota_j(\widetilde{\bv})-1}{2^{j+1}}\right) - F_f^{-1}\left(  \frac{2 \iota_j(\bv^+)-1}{2^{j+1}}\right),
\eeq
and
\beq
\Sigma_j(\bv^+, \widetilde{\bv}) =  F_f^{-1}\left(  \frac{2 \iota_j(\widetilde{\bv})-1}{2^{j+1}}\right) + F_f^{-1}\left(  \frac{2 \iota_j(\bv^+)-1}{2^{j+1}}\right).
\eeq

Now consider the transmitted symbol sequence $\bv_1^{n+d-1}$. Since the CCM scheme satisfies the forward ordering property and $F_f^{-1}$ is a nondecreasing function, it is easy to verify that, when $s_n(\widetilde{\bv}) > s_n(\bv^{+})$, $\iota_j(\bv_1^{n+d-1}) \leq \iota_j(\bv^+) \leq \iota_j(\widetilde{\bv})$ for $i=1,\dots,n-1$ and $\iota_j(\bv_1^{n+d-1}) \leq \iota_j(\bv^+) < \iota_j(\widetilde{\bv})$ for $i=n,\dots,n+d-1$. As a consequence, $\zeta\left(\sv(\bv_1^{n+d-1}), \bv^+, \widetilde{\bv}\right) < 0$. Moreover, the distance between $\sv(\bv_1^{n+d-1})$ and the hyperplane $\pi(\bv^+,\widetilde{\bv})$ is given by
\begin{eqnarray}
d\left(\sv(\bv_1^{n+d-1}), \pi(\bv^+,\widetilde{\bv}) \right) & = & \frac{\left| \zeta\left(\sv(\bv_1^{n+d-1}), \bv^+, \widetilde{\bv}\right)\right|}{2\nu(\bv^+,\widetilde{\bv})} \non
& = & \frac{\sum_{j=1}^{n+d-1} \Gamma_j^2 \Delta_j(\bv_1^{n+d-1},\bv^+ )\Delta_j(\bv^+, \widetilde{\bv}) }{\nu(\bv^+,\widetilde{\bv})} + \frac1{2}\nu(\bv^+,\widetilde{\bv}) \non
& \geq & \frac1{2}\nu(\bv^+,\widetilde{\bv}), 
\label{eq:dist_lb} 
\end{eqnarray}
where we have defined
\beq
\nu(\bv^+,\widetilde{\bv}) = \sqrt{\sum_{j=1}^{n+d-1} \Gamma_j^2 \left(\Delta_j(\bv^+, \widetilde{\bv})\right)^2}.
\eeq
We can lower-bound the RHS of \eqref{eq:dist_lb} by minimizing $\Delta_j(\bv^+, \widetilde{\bv})$ over the possible values of $\widetilde{\bv}$. It turns out that
\beq
\min_{\widetilde{\bv}} \Delta_j(\bv^+, \widetilde{\bv}) = \left\{
\begin{array}{cc}
0, & j < n \\
F_f^{-1}\left(  \frac{2 \iota_j(\bv^+)+1}{2^{j+1}}\right) - F_f^{-1}\left(  \frac{2 \iota_j(\bv^+)-1}{2^{j+1}}\right), & j \geq n
\end{array}
\right. .
\eeq

It is easy to see that, whatever is the bit mapping, $\iota_j(\bv^+) = 2^{j-n} \iota_n(\bv_1^{n})$, so we can easily obtain that 
\beq \label{eq:dist_lb_2}
d\left(\sv(\bv_1^{n+d-1}), \pi(\bv^+,\widetilde{\bv}) \right) \geq \bar{\rho}\left(n,d, \iota_n(\bv_1^{n}) \right).
\eeq
As a consequence of \eqref{eq:dist_lb_2}, all points $\rv_1^{n+d-1}$ in the hypersphere centered on $\sv(\bv_1^{n+d-1})$ with radius $\bar{\rho}\left(n,d, \iota_n(\bv_1^{n}) \right)$ satisfy $\zeta\left(\rv_1^{n+d-1}, \bv^+, \widetilde{\bv}\right) < 0$ and are thus closer to $\sv(\bv^+)$ than to $\sv(\widetilde{\bv})$. 

Exactly the same procedure can be followed to show that all points $\rv_1^{n+d-1}$ in the hypersphere centered on $\sv(\bv_1^{n+d-1})$ with radius $\bar{\rho}\left(n,d, \iota_n(\bv_1^{n})-1 \right)$ satisfy $\zeta\left(\rv_1^{n+d-1}, \bv^-, \check{\bv}\right) < 0$, when $s_n(\check{\bv}) < s_n(\bv^{-})$, and are thus closer to $\sv(\bv^-)$ than to $\sv(\check{\bv})$. 
Thus, the proof of the lemma is complete.
\else
\fi

%
%

\ifCLASSOPTIONcaptionsoff
  \newpage
\fi

\end{document}